\documentclass[pra,amsmath,amssymb,showpacs]{revtex4}
\usepackage{epsfig}

\begin{document}

\title{Symmetries and Geometries of Qubits, and their Uses}


\date{\today}

\author{A. R. P. Rau\footnote{e-mail: arau@phys.lsu.edu}}

\affiliation{Department of Physics and Astronomy, Louisiana State University, Baton Rouge, Louisiana 70803, USA}

\begin{abstract}
{\bf Abstract}
The symmetry SU(2) and its geometric Bloch Sphere rendering are familiar for a qubit (spin-1/2) but extension of symmetries and geometries have been investigated far less for multiple qubits, even just a pair of them, that are central to quantum information. In the last two decades, two different approaches with independent starting points and motivations have come together for this purpose. One was to develop the unitary time evolution of two or more qubits for studying quantum correlations, exploiting the relevant Lie algebras and especially sub-algebras of the Hamiltonians involved, and arriving at connections to finite projective geometries and combinatorial designs. Independently, geometers studying projective ring lines and associated finite geometries have come to parallel conclusions. This review brings together both the Lie algebraic and group representation perspective of quantum physics and the geometric algebraic one, along with connections to complex quaternions. Together, all this may be seen as further development of Felix Klein's Erlangen Program for symmetries and geometries. In particular, the fifteen generators of the continuous SU(4) Lie group for two-qubits can be placed in one-to-one correspondence with finite projective geometries, combinatorial Steiner designs, and finite quaternionic groups. The very different perspectives may provide further insight into problems in quantum information. Extensions are considered for multiple qubits and higher spin or higher dimensional qudits.    

\end{abstract}
\maketitle

    



\section{Introduction}
This article deals with symmetry aspects of systems in quantum information \cite{ref1}. The unitary group SU(2) symmetry of a single two-level system, a ``qubit," is well-known and exploited widely throughout physics, not just in quantum information. It is taught in the very first courses on quantum spin-1/2. The so-called Bloch sphere representation \cite{ref2, ref3} has been used from the earliest days of nuclear magnetic resonance (nmr). Its mapping of the unitary transformations of complex wave functions onto rotations of a classical unit vector on an ordinary globe's sphere $S^2$ pervades the very language of nmr and its applications in chemistry and medicine. It provides even the intuition for transformations between quantum states in terms of axes, and angles of rotation about them, to achieve a desired end. ``Apply a $\pi$-rotation (`spin-flip') about some axis followed by a different rotation around a second axis" forms the very lore of the subject \cite{ref4}. In contrast, the similar SU(4) that pertains to a two-qubit system, or higher dimensional extensions and their corresponding geometrical pictures, have been little used in the field of quantum information even though one might have expected it given the central role that two or more entangled spins play in quantum computing, quantum cryptography, and other sub-areas of the field. This is also surprising since quantum physics and its various developments in atomic, nuclear, condensed matter, and particle physics have repeatedly shown the importance, even necessity, of using symmetry ideas both for understanding and insight and as technical aids to facilitate calculation. Further, already in classical physics, rotations and associated symmetries of orthogonal groups SO($d$) were studied from the beginning in physics and later for quantum angular momentum, both orbital and spin. And quantum physics also introduced the unitary groups SU($d$) as natural objects for studying time evolution of quantum states.

This review will deal with symmetry considerations of the operators and states involved to permit understanding and convenient calculation of quantum correlations such as entanglement \cite{ref1}, quantum discord \cite{ref5,ref6}, and a variety of others. We will consider two or more qubits and also higher dimensional ``qudits"  of dimension $d$ larger than two. These involve higher unitary groups such as SU($2^q$) for $q$ qubits or SU($d$) for a single qudit. Whereas the SU(2) group, and corresponding su(2) algebra (we follow the convention of using upper case for Lie groups and lower case for Lie algebras), involve three generators and parameters, the higher groups deal of course with many more. The SU(4) for two qubits has fifteen and any general SU($d$) has $(d^2-1)$ generators. In part, these larger numbers may have deterred their use. Their consideration nevertheless proves useful, especially because smaller sub-groups often come into play in the very physics and symmetries involved in the system of interest. This will be a recurring theme in this review. As an example, many sub-groups of SU(4) have been shown to represent some of the logic gates and Hamiltonians arising in coupled two-qubit systems (or in more general 4-level systems of atomic and molecular physics and quantum optics \cite{ref7,ref8}). And one such sub-group, SU(2) $\times$ U(1) $\times$ SU(2), is also the symmetry of the 7-parameter $X$-states \cite{ref9} that were previously defined and discussed as useful for a variety of qubit-qubit problems before this underlying symmetry was recognized \cite{ref10}.

A wider context for our discussion is provided by the famous 1872 Erlangen Program of Felix Klein \cite{ref11}. Instead of the centuries-old view of geometry as a set of axioms defining points, lines, triangles, circles, and of theorems relating to their properties, Klein redefined geometry in terms of the study of symmetries and associated transformations, each such set then defining a geometry. Euclidean geometry is but one, the one that stems from Euclidean transformations in a plane. But, there can be an infinity of others depending on the symmetries and transformations initially specified. This re-orientation and focus on geometries in the plural, each associated with a set of symmetries, has had a profound effect on mathematics and physics ever since. Coupled with Emmy Noether's theorem that associates a continuous symmetry with a conservation law, an invariant of that symmetry, the most fundamental laws of physics, namely, the conservation laws of linear and angular momentum, energy, charge, etc., place symmetries at the heart of physics \cite{ref2,ref3,ref12}. 

Together with Klein's contemporary colleague and friend Sophus Lie \cite{ref12a}, who developed the subject of continuous symmetry groups such as SO($d$) and SU($d$), their influence on modern physics cannot be stressed enough. Havel and Doran \cite{ref13} have a striking diagram on this historical influence. Reproduced here as Fig. 1, they classify subsequent work in three ladders. The first is a study of invariants while expressing geometric relations, subsumed under a general study of algebraic curves and surfaces in the area of mathematics called algebraic geometry. A second ladder, initiated by Grassmann and tying to quaternions that were invented by Hamilton even before the Klein Program, has come to be known as geometric algebra. Closely linked to Clifford algebras, it is the study of tensors and spinors of various ranks, differential forms, etc. Its value has been recognized in recent times in the work of Baylis, Sobcyzk, Doran, Lasenby, Hestenes, and others. The last named author, in particular, has through several works and textbooks, shown how classical and quantum mechanics, non-relativistic and relativistic, can be described in geometric algebra to great advantage, even extending to pedagogical benefit, compared to alternative treatments that have otherwise become standard \cite{ref14,ref15,ref16}. The third ladder is group representation theory originated by Frobenius, and then mainly the work of Lie and Engel in the area of mathematics called Lie groups and Lie algebras \cite{ref17}. Because of the wide role for orthogonal SO and unitary SU groups, physics students are most familiar with this third ladder and the matrix representation that is most commonly used.

\begin{figure}
\centering
\scalebox{1.5}{\includegraphics[width=2.7in]{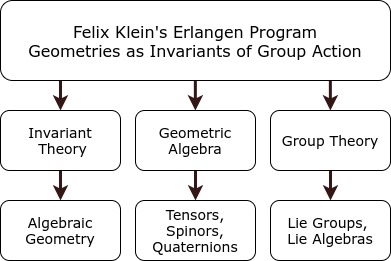}}
\caption{Felix Klein's Erlangen Program and its subsequent developments along three ladders, adapted from \cite{ref13}. Klein initiated the view of geometries as characteristic of symmetries of transformations and their symmetry groups. The study of invariants and polynomials in them led to the study of algebraic curves and surfaces and what is termed algebraic geometry. Tensors, spinors, quaternions, and Clifford algebras, and the study of differential forms came to be known as geometric algebra. Its passionate followers advocate adopting it over what physicists have come to call ``vector algebra," which is not really an algebra, that has prevailed in physics. A third chain was the study of group representations, of both discrete groups and continuous Lie groups and Lie algebras, matrix representations of them having come to dominate contemporary physics.}
\end{figure}

Although symmetries and transformations are important also in classical physics, they became even more crucial from the very beginnings of quantum physics and its applications in atomic, condensed matter, nuclear, and particle physics. Most physicists are, therefore, familiar with SU(2), SU(3), SU(4), SO(3), SO(4), etc., while not as much with geometric algebra, quaternions, Grassmann manifolds, and such. But, in the last 20 years, different lines of work in quantum information and especially the study of a pair of qubits has pointed to further inter-connections between the topics described in the previous paragraph; in particular, connections across the different ladders in Fig. 1. These links between  continuous Lie groups and algebras and finite discrete groups of one or more quaternions, and further connections to finite projective geometries (another field pioneered by Klein and contemporaries, notably Gino Fano) and so-called balanced incomplete block designs \cite{ref18,ref19} (in an area called design theory \cite{ref20}) will be discussed in this review. It is remarkable that the fifteen generators of the qubit-qubit SU(4) symmetry provide links between these disparate fields. The same number of generators in re-arranged combinations describe SO(6), the group of six-dimensional rotations, and its closely allied non-compact groups SO(4,2) of the hydrogen atom in non-relativistic quantum mechanics \cite{ref2,ref21}. Also, there are fifteen Dirac matrices that occur throughout relativistic field theories and particle physics \cite{ref22}, and fifteen points form the important finite projective geometry PG(3,2). The finite group of complex quaternions also has the same number of elements. Thus, our discussion of these disparate elements brings many cross-connections between the three ladders listed in the previous paragraph, providing additional insight into what seem to be widely different topics. 

This Review is broadly in two parts. Section II consists the first, dealing with the time evolution operator of multiple qubits in quantum information, derived on the basis of a Lie algebra of generators that close under commutation. The second, in Sections III and IV, deals with the various sub-algebras that may be involved and correspondences to other elements of Fig. 1 such as geometric algebra, discrete groups of quaternions, finite projective geometries, and combinatorial designs.      
   
\section{Unitary evolution operator}
As more generally throughout quantum physics, the unitary operator $U(t)$ of time evolution for a Hermitian Hamiltonian $H(t)$ plays a central role. (Even more general extensions to non-Hermitian Hamiltonians and Lindblad type master equations for handling decoherence and dissipation are possible, through embedding the elements of a $n \times n$ density matrix in an $(n^2-1)$ dimensional space \cite{ref23}, but our discussion will confine to Hermitian and unitary language.) The basic equations of motion, and the time development of wave functions, density matrices, and operators are governed by this $U(t)$. The algebraic and geometric description of this unitary operator is, therefore, central to quantum physics and quantum information science. This review's focus is on $U(t)$ which took on even more importance in quantum information following an important paper by Luo \cite{ref24} that set the course for handling correlations such as quantum discord of a composite system AB that call for all possible measurements on one of the sub-systems. The logic is that such local measurements on A or B alone cannot change quantum correlations but can provide all classical correlations. They can then be separated from the total correlations in AB to leave behind what must be quantum correlations. 

To give concrete meaning to what is meant by {\it all} possible measurements, either in a theoretical calculation or as an operational experimental procedure, Luo \cite{ref24} considered for a qubit the pair of Stern-Gerlach or von Neumann projections with respect to some $z$-axis, $\Pi_{\pm}$, and then subjected them to a general unitary transformation,

\begin{equation}
A_i = U_i \Pi_{\pm} U_i ^{\dagger}.
\label{eqn1}
\end{equation}
This provides a well-defined procedure to handle all possible measurements and indeed resonates intuitively with the physics of a charged spin-1/2 particle. A measurement on it constitutes a Stern-Gerlach one with two possible outcomes: the antipodal points on the Bloch sphere. One then rotates that axis of orientation of the anisotropic magnet through all positions in three-dimensional space, thus exhausting all possibilities for measuring the qubit. This can of course be generalized to other spins or qudits with possibly more general POVM (positive operator valued measure \cite{ref25}) than a von Neumann projector in between but again using a general unitary $U(t)$ for that dimension in Eq.~(\ref{eqn1}). This points to the need for such unitary evolution operators for multi-qudit systems in order to understand correlations or to construct logic gates in quantum information.  

For the SU(2) of a qubit, a general unitary transformation is unambiguous, and well known:

\begin{equation}
U_i = t I +i\vec{y} \cdot \vec{\sigma}, \, \, t^2+{\vec{y}}^2 = 1,
\label{eqn2}
\end{equation}
and these three parameters then describe all the measurements on A.   
This basic procedure of Luo \cite{ref24}, initially for a very limited subset of qubit-qubit states, was later adapted \cite{ref26,ref27} for a larger class of density matrices called $X$-states \cite{ref9,ref10} of the form,

\begin{equation}
\rho =  \left( 
\begin{array}{cccc}
\rho_{11} & 0 & 0 & \rho_{14} \\ 
0 & \rho_{22} & \rho_{23} & 0 \\ 
0 & \rho_{32} & \rho_{33} & 0 \\
\rho_{41} & 0 & 0 & \rho_{44}
\end{array}
\right).      
\label{eqn3}
\end{equation}

They were at first so-named for their visual appearance, non-zero entries standing only along the diagonal and anti-diagonal in such a 4 $\times$ 4 matrix representation in the canonical basis. With three real diagonal elements and two complex off-diagonal ones, this is a 7-parameter set. (Local unitary transformations can reduce this to 5 real ones by removing the phase angles of the complex elements through unitary rotations \cite{ref28}.) While smaller than the full 15 parameters of the most general qubit-qubit density matrix, many calculations of entanglement and other properties, and their evolution under unitary or dissipative processes, can be easily carried out for such states which make them appealing objects for study. Many specific states of interest, such as the maximally entangled Bell states \cite{ref1} and `Werner' states \cite{ref29}, are a sub-class of $X$-states, lending further importance to their study. The fewer parameters in an $X$-state do not restrict the range of physical phenomena investigated. A large variety of qubit-qubit physics can thus be discussed through $X$-states. At the same time, the restriction in the number of parameters allows for ready calculation accounting for their popular use. The symmetry group and algebra of $X$-states will be discussed in detail in Sec. III A. 

Even though SU(2) has three parameters as does $U(t)$ in Eq.~(\ref{eqn2}), it turns out that only two are of interest, and can be identified with the two angles on the Bloch sphere. This reduction is clearest in the symmetry decomposition of SU(2) as the base S$^2$ and a one-dimensional ``fiber" U(1) or pure phase. That latter element commutes past the projector in Eq.~(\ref{eqn1}) so as to cancel itself out in $U$ and $U^{\dagger}$, to leave only the two Bloch angles of S$^2$. This has been formalized in a simple prescription in \cite{ref27} applicable for calculations such as quantum discord in a general density matrix of AB for any dimension of B so long as A is a qubit. This prescription also removes the restriction to $X$-states in the previous paragraph and can handle all qubit-qubit density matrices. The symmetry decomposition of the SU(2) $U(t)$ will be taken up in Sec. II A. It leads suggestively to a similar treatment of any SU($N$) in Sec. II B, thereby providing a compact and simple procedure for constructing $U(t)$ for any dimension.
Further, for $X$-states, the $\theta$ Bloch angle (latitude) describing $U(t)$ suffices with $\phi$ (longitude) unnecessary. Studies \cite{ref27,ref30} have found that in over 99\% of tens of thousands of randomly chosen density matrices, the extremum that gives the quantum discord is reached at the extreme angle $\theta =\pi/2$. The initial prescription \cite{ref26} in terms of this seems to leave a very small worst case error \cite{ref31}. 

\subsection{Derivation of evolution operator for a qubit}

Solving the Schr\"{o}dinger equation for the evolution operator, $i dU(t)/dt=H(t)U(t)$, $U(0)= I$, for a qubit is straightforward. We will set $\hbar =1$. A non-zero trace of $H$ can first be filtered out as a phase factor and, since the rest can be cast in terms of the three Pauli spin operators, correspondingly $U(t)$ has three exponential factors with each Pauli spinor multiplying a time-dependent coefficient in the exponent. Conventionally, one would use the three Cartesian Pauli spinors, the coefficients being then real, Euler angles of rotation. They obey a system of coupled, first-order in $t$, differential equations familiar from Euler equations for rigid-body rotations in classical mechanics \cite{ref32}. (This connection also points to differing terminologies because, for many purposes, one could refer to so(3) instead of su(2) or SU(2) but we will use the unitary language in this review; however, some of the mathematics literature uses so(3) instead.) But these Euler equations are highly nonlinear, involving sines and cosines of the coefficients. Instead, and as also the one convenient for our later generalization to higher SU($N$), we \cite{ref33,ref34,ref34a} use the step-up/down combinations $\sigma_{\pm} \equiv \sigma_x \pm i \sigma_y$ that lead to simpler equations and interpretations. 

\begin{figure}
\centering
\scalebox{2.5}{\includegraphics[width=2.7in]{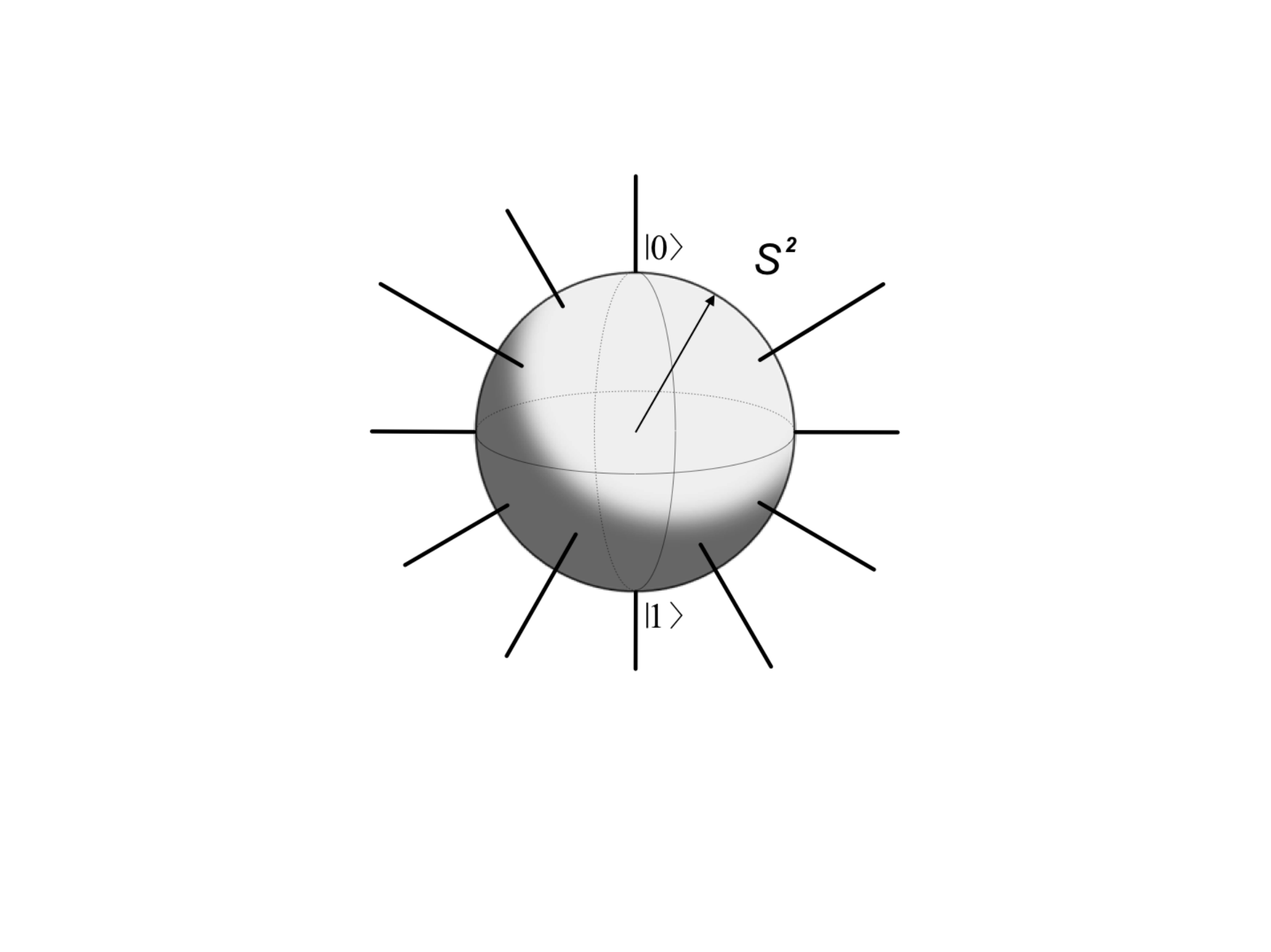}}
\vspace{-1.2in}
\caption{The fiber bundle for SU(2), with the Bloch sphere as a base manifold, and spikes at each point on it representing a U(1) phase. The three parameters defining a point on the sphere in terms of two Bloch angles, and a value for the phase at that point, provide the complete description of the dynamics of a spin-1/2 system. That dynamics amounts to rotations of the unit Bloch vector shown. From \cite{ref34}.}
\end{figure}   

Along with $\sigma_z$, this alternative triplet choice of Pauli matrices also spans the complete algebra, so that a full solution for $U(t)$ takes the form of a product of three exponential factors,   

\begin{equation}
U(t) = e^{z(t) \sigma_{+}/2} e^{w^{*}(t) \sigma_{-}/2} e^{- i\mu(t) \sigma_z/2}.
\label{eqn4}
\end{equation}
The complex quantities $z$, $w$, and $\mu$ are classical functions, vanishing at $t=0$. The solution is by construction \cite{ref33}. Plugging Eq.~(\ref{eqn4}) into the evolution equation and re-arranging operators through the Baker-Campbell-Hausdorff identity \cite{ref35} is all that is required to get the defining equations for $z$, $w$, and $\mu$, and to get the relations between them that guarantee that the overall $U(t)$  is unitary even if individual factors in Eq.~(\ref{eqn4}) no longer are as in the Cartesian decomposition. Also, those three relations reduce the three complex coefficients now involved again to three real independent parameters. With $w$ essentially the complex conjugate of $z$, the three linearly independent quantities may be taken to be the real and imaginary parts of $z$ and Re $\mu$. The last is determined by quadrature involving $z$ while $z$ itself is solved from a self-contained Riccati equation \cite{ref33,ref34,ref34a}. 

The complex quantity $z$ may then be inverse stereographically projected onto the ``Bloch sphere" \cite{ref2} through defining a unit three-dimensional vector $\vec{m}$ \cite{ref34}. The nonlinear Riccati equation for $z$ then becomes the linear Bloch equation, $d \vec{m}/dt =-2\vec{B} \times \vec{m}$. However, $z$, $\vec{m}$, and the Bloch sphere account for only two of the three parameters of the full SU(2) problem, the third being the phase parameter Re $\mu$. Thus, the full (local) description geometrically is a ``spiked unit sphere" as shown in Fig. 2, with the spike at each point on the sphere representing a $U(1)$ phase. This is what mathematicians call the ``fiber bundle" [SU(2)/U(1)] $\times$ U(1), with [SU(2)/U(1)] $ \sim S^2$ the ``base manifold" and the U(1) phase the one-dimensional ``fiber" \cite{ref36}. The evolution operator $U(t)$ in Eq.~(\ref{eqn4}) can be pictured schematically as in Fig. 3. Its very structure, with the first two factors in Eq.~(\ref{eqn4}) triangular and the third diagonal, suggests easy generalization to be considered next in Sec. II B.

\begin{figure}
\centering
\scalebox{2.5}{\includegraphics[width=2.7in]{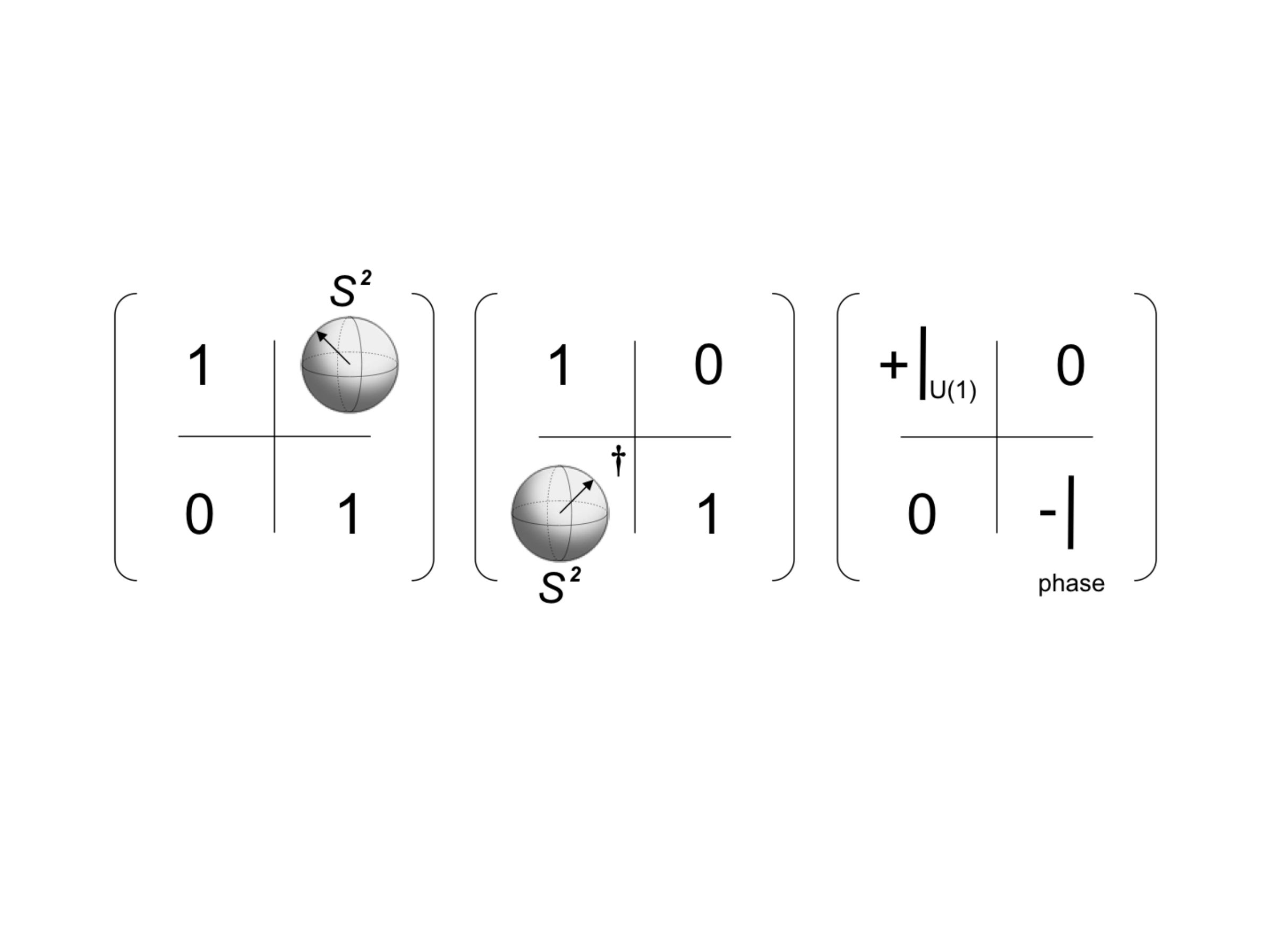}}
\vspace{-1.2in}
\caption{Structure of the $2 \times 2$ matrix evolution operator in Eq.~(\ref{eqn4}) for SU(2). The three factors involve, respectively, the Pauli spinors $\sigma_{+}$, $\sigma_{-}$, and $\sigma_z$. A single complex number $z$ provides the first two matrices, shown schematically as the Bloch sphere obtained through inverse stereographic projection of $z$. The third factor is a diagonal matrix, defined through a single number or phase depicted as a line or spike, which enters with opposite signs in the two entries. The complete fiber bundle, SU(2) : $S^2$ $\times$ U(1), in Fig. 2 may be viewed in the above factorized form. The same form of three factors can be generalized, as discussed later in Fig. 4 in Sec. II B, to any SU($N$), with the first two factors providing the base manifold and the third the fiber. From \cite{ref34}.}  
\end{figure}

This unitary integration procedure has as its its key feature, and as the only algebraic manipulation needed, the Baker-Campbell-Hausdorff identity \cite{ref35} that involves a sequence of successive commutators of the operators that occur in the problem. It is for this reason that Lie algebras and Lie groups arose and fit naturally into quantum physics applications. It also means that unitary integration seems to have been introduced independently several times even if not named as such. The earliest occurrence may be in \cite{ref37} so that it may be referred to as the Wei-Norman method. It is related to but different from and more convenient than the Magnus expansion \cite{ref37a}, and has had a revival since the mid-1980s for su(2), su(1,1), and also quantum spin-1 cases by many different groups \cite{ref38,ref39,ref40}. For problems involving two qubits and the algebra su(4), again different applications were made, independently, in the last twenty years for Cartan decompositions of su(4) in quantum control \cite{ref41,ref42} and more generally \cite{ref43}. A decomposition of $U(t)$ for su(4) into factors of local unitaries of individual qubits and a 3-parameter diagonal unitary matrix was given independently in \cite{ref42,ref43a} and \cite{ref43b}. Related work on generating entanglement dynamics and the minimum number of unitaries required is in \cite{ref43c}.

With 15 generators and their commutators involved (and 8 for su(3) of a single spin-1), products of that many exponentials in Eq.~(\ref{eqn4}) become unwieldy. Symmetries present that restricted to the smaller number of generators of sub-algebras, especially in a nmr problem using just 7, helped practical implementation \cite{ref7}. We will consider in the next section the derivation of $U(t)$ that can handle this and then in Sec. III consider some of the main sub-algebras involved of su(4) and their associated physical systems.   

\subsection{Derivation of evolution operator for SU($N$)}

The above construction of $U(t)$ in the form of a three-term product as in Eq.~(\ref{eqn4}) and Fig. 3 can be carried over to any $N$-level or SU($N$) problem. Thereby, the form and simplicity of the $N=2$ case can be extended to arbitrarily large $N$. The final result is very simply stated based on symmetry patterns alone. View the $N$-dimensional Hamiltonian ${\bf H}^{(N)}$ as $2 \times 2$ blocks,

\begin{equation}
{\bf H}^{(N)} =\left(
\begin{array}{cc}
{\bf H}^{(N-n)} & {\bf V} \\
{\bf V}^{\dagger} & {\bf H}^{(n)}
\end{array}
\right),
\label{eqn5}
\end{equation}
with $n, 1\leq n <N$, an arbitrary choice. The diagonal blocks are square matrices while the off-diagonal ${\bf V}$ is $(N-n) \times n$ and ${\bf V}^{\dagger}$ is $n \times (N-n)$. These latter are taken as Hermitian adjoints and ${\bf H}^{(N)}$ as traceless although much of our construction applies more generally \cite{ref23,ref34}. 

\begin{figure}
\centering
\scalebox{2.5}{\includegraphics[width=2.7in]{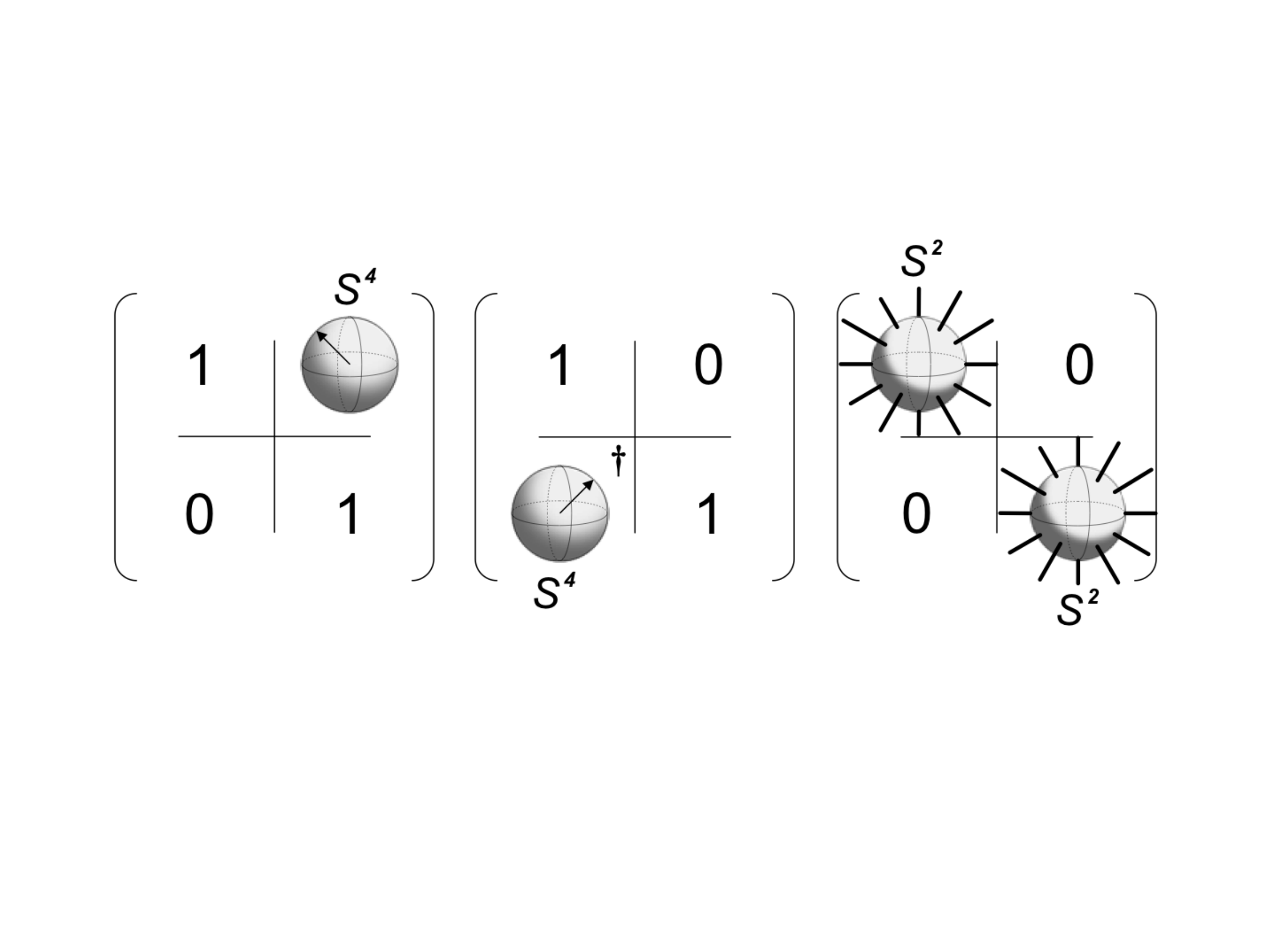}}
\vspace{-1.2in}
\caption{Analogous to Fig.\ 3, schematic of the evolution operator for general su($N$) as three factors but now of block matrices, the last of diagonal blocks of lower dimension. For concreteness, the so(5) sub-algebra of su(4) is illustrated when $z$ consists of four real parameters, shown as inverse stereographically rendered by a four-sphere $S^4$, and the third factor has two su(2) blocks along the diagonal. This is shown geometrically in Fig. 6.}
\end{figure}

With such a $2 \times 2$ block view of the $N$-dimensional $H$, a solution for $U(t)$ can be written in close analogy to Eq.~(\ref{eqn4}) as a product of three $2 \times 2$ block matrices, the first two having one non-zero off-diagonal block while the last is block-diagonal. This structure of two idempotent (the multiplication of two such giving a similar matrix with a zero off-diagonal block) factors and a third that is block diagonal was a result of the choice of step-up/down rather than Cartesian operators in Eq.~(\ref{eqn4}) and proves crucial. Rectangular matrices, ${\bf z}$ and ${\bf w}^{\dagger}$, now stand in place of complex numbers $z$ and $w$, respectively, in Eq.~(\ref{eqn4}). This key step is illustrated in Fig. 4 in direct analogy to Fig. 3, but now with block matrices as the entries in the 2 $\times$ 2 structure. First, one solves the equation satisfied by ${\bf z}$, now a matrix Riccati equation \cite{ref44}. For unitary problems, the matrix ${\bf w}$ is simply related to ${\bf z}$, again as in the $N=2$ case \cite{ref34}. Also from ${\bf z}$, effective Hamiltonians are constructed for the two diagonal blocks of the third factor in $U(t)$ for subsequent handling as smaller SU($N-n$) and SU($n$) problems. We refer to \cite{ref34} for details. For any $N$, and choice of $n$, the procedure can thus be iterated down to a final SU(2). One can describe this construction as following in spirit Schwinger's method of building higher angular momentum $j$ representations of SU(2) as products of the ``fundamental" $j=1/2$ \cite{ref2,ref3}. But, instead of different representations of the same SU(2) that is involved there, this time we construct representations of different SU($N$) in terms of a succession of three-factor products of block matrices of the same form as the three Pauli factors in SU(2)'s Eq.~(\ref{eqn4}).
 
Typically, $n$ will be chosen to be 1 or 2. In the former case, ${\bf z}$ is a vector of $(N-1)$ complex $z_i$, the lower diagonal block in the effective Hamiltonian a single element, and its corresponding element in $U(t)$ a phase. This generalizes the fiber bundle description for SU(2) to SU($N$) with a base manifold SU($N$)/[(SU($N-1$) $\times $ U(1)]. Iteratively, one can reduce from $N$ to a lower value, extracting a U(1) phase at each step \cite{ref34,ref34a}. This is called the flag manifold. The construction for SU(3) is given in \cite{ref45}. The case $n=2$ is also interesting and useful, this time the lower block in $H$ being a $2 \times 2$ matrix. Using Pauli algebra, again the Hamiltonians and unitary matrices at each step can be explicitly worked out \cite{ref34}. We will return to this in Secs. III B and III C after first considering sub-group symmetries that may apply in a high $N$ case.

\section{Sub-group symmetries and sub-algebras}

The previous section constructed $U(t)$ for any SU($N$) no matter how large $N$ is. There is an explosive growth of $N$ for multiple qubits, with $N=2^q$ for $q$-qubits, even larger for multiple qudits of higher dimension. However, physical situations often introduce further symmetries that limit to a smaller number which can be handled more easily. Already, as noted, even for SU(4) of $q=2$ with fifteen operators and parameters in general, Hamiltonians and states may involve fewer so that identifying a corresponding sub-group to which they belong can be useful. Many SU(2) and SU(2) $\times$ SU(2) are trivial examples of such sub-groups of SU(4). The latter would pertain to two completely independent spins with no coupling between them. In that case, $U(t)$ need not be written as a product of fifteen exponentials but more simply as two factors of the form of Eq.~(\ref{eqn4}) for each qubit or SU(2). These six operators and the unit operator themselves close under commutation, further dividing into two independent sets. The other nine need not be invoked at all. The unit operator is the only one that commutes with all in this sub-group, the only ``center" in the language of group theory. 

A less trivial, but very important, example is a sub-group SU(2) $\times$ U(1) $\times$ SU(2) which involves seven operators and parameters in a Hamiltonian with this sub-group symmetry. The U(1) is a single, but non-trivial, operator that also is a center, commuting with all six others which themselves can be arranged as two independent SU(2) or sets of three. In quantum error correction, it is referred to as the ``stabilizer" \cite{ref45a}. In this case again, $U(t)$ splits into two independent factors as in Eq.~(\ref{eqn4}) and an additional exponential in that U(1) element. Concrete examples occur in logic gates and Hamiltonians in quantum information \cite{ref7}. Two independent spiked Bloch spheres as in Fig. 1 and the previous section, along with a U(1) fiber element linking them, give a geometrical rendering of such a sub-group symmetry. The decomposition into the sub-algebra of the previous paragraph is referred to as so(3) $\oplus$ so(3) in Eqs.(45,46) of a general mathematical study \cite{ref13} which also noted a seven-dimensional so(2) $\oplus$ so(3) $\oplus$ so(3) in its Sec. V. Quantum physics applications pointed to its importance in a variety of problems \cite{ref7,ref34}. 

Other sub-groups of SU(4) include, of course, several SU(3) of sets of eight generators as well as SO(5), the rotation or orthogonal group in five dimensions with 10 generators. Again, many instances occur in quantum optics, quantum information, atomic and molecular physics \cite{ref34}. To identify all such sub-groups systematically, a table of commutators of all fifteen operators of SU(4) shown in Table I proves useful \cite{ref8,ref34,ref46}. It follows immediately from inspection, for instance, that every row or column of this table has seven zeroes which means that every one of the fifteen can play the role of that non-trivial center in a SU(2) $\times$ U(1) $\times$ SU(2) sub-group. (Other sets that close under commutation, of eight or ten, can also be seen in Table I to represent SU(3) and SO(5) symmetry, respectively, to be discussed below in Sec. III B.) 

Different notations have proved useful, a sequential set $O_i, i=1, 2, \ldots 16$ \cite{ref7,ref34,ref46} applicable to any four-level system, or direct products of two sets of Pauli operators when there is a two-qubit origin: $( I, I \otimes \sigma_i, \tau_i \otimes I, \tau_i \otimes \sigma_j)$. Two different symbols $\sigma$ and $\tau$ prove convenient for two independent spin-1/2 but for easier generalization to more qubits, an upper index $\vec{\sigma}^{(i)}$ serves better \cite{ref46}. When dealing with the three components, the natural short-hand notation of $(X, Y, Z)$ proves convenient. Thus, $XZ$ denotes $\tau_x \sigma_z$ or $\sigma^{(2)}_x \sigma^{(1)}_z$ or $O_{11}$, while $IX$ is $\sigma_x$ and $ZI$ is $\tau_z$, or $O_5$ and $O_3$, respectively. Table II gives the complete list along with a correspondence to Dirac gamma matrices \cite{ref22} used in relativistic quantum field theories. Yet another is a convenient 4-binary labelling that we will take up in Sec. III E. And a mapping onto complex quaternions and their finite groups along with a different binary labelling will be discussed in Sec. III F. Yet another labelling in terms of bivectors $G_{ij}$ to be discussed in Sec. IV is also shown. The Dirac gamma matrices constitute four four-vectors $\gamma_{\mu}$, $\mu =1-4$, denoted V and obeying anti-commutation relations, six anti-symmetric products of two of them denoted T(ensor) as $\sigma_{\mu \nu} = -\frac{i}{2} \gamma_{\mu} \gamma_{\nu}$, a P(seudo-scalar) $\gamma_5$ that is the product of all four gamma matrices, and four pseudo-vectors A given by $i\gamma_5 \gamma_{\mu}$ \cite{ref22}. 

\begin{table*}
\begin{center}
\begin{tabular}{|c||c|c|c|c|c|c|c|c|c|c|c|c|c|c|c|c|}

\hline
$O_X $&$O_2 $&$O_3 $&$O_4  $&$O_5  $&$O_6 $&$O_7 $&$O_8 $&$O_9 $&$O_{10} $&$O_{11} $&$O_{12}$&$O_{13}$&$O_{14} $&$O_{15} $&$ O_{16}  $\\ \hline \hline
$O_2 $&$0   $&$0    $&$0    $&$iO_6  $&$-iO_5  $&$iO_8 $&$-iO_7$&$0     $&$0     $&$0     $&$0      $&$iO_{16} $&$-iO_{15}$&$iO_{14}  $&$-iO_{13}  $\\
\hline
$O_3 $&$0   $&$0    $&$ 0    $&$0     $&$0      $&$0     $&$ 0 $&$iO_{10} $&$-iO_9 $&$iO_{12} $&$-iO_{11} $&$iO_{15} $&$-iO_{16}  $&$-iO_{13} $&$ iO_{14} $\\ \hline
$O_4 $&$0   $&$0    $&$0    $&$iO_8  $&$-iO_7  $&$\frac{i}{4}O_6$&$-\frac{i}{4}O_5$&$iO_{12} $&$-iO_{11}$&$\frac{i}{4}O_{10}$&$-\frac{i}{4}O_9$ &$0$&$0$&$0$&$0$\\
\hline
$O_5 $&$-iO_6$&$ 0 $&$ -iO_8 $&$ 0 $&$ iO_2 $&$ 0$&$ iO_4 $&$ 0 $&$ 0 $&$-iO_{16}$&$-iO_{14}$&$ 0 $&$iO_{12}$&$0$&$iO_{11}$\\
\hline
$O_6 $&$iO_5 $&$ 0 $&$iO_7$&$-iO_2$&$0$&$ -iO_4 $&$0$&$0$&$ 0 $&$ iO_{13} $&$iO_{15}$&$-iO_{11}$&$0$&$-iO_{12}$&$ 0 $\\
\hline
$O_7 $&$-iO_8 $&$0$&$-\frac{i}{4}O_6$&$0$&$iO_4$&$0$&$\frac{i}{4}O_2$&$iO_{15}$&$-iO_{13}$&$0$&$0$&$\frac{i}{4}O_{10}$&$0$&$-\frac{i}{4}O_9$&$0$\\ \hline
$O_8 $&$iO_7 $&$ 0 $&$\frac{i}{4}O_5$&$-iO_4$&$ 0 $&$-\frac{i}{4}O_2$&$0$&$iO_{14}$&$-iO_{16}$&$0$&$0$&$0$&$-\frac{i}{4}O_9$&$0$&$\frac{i}{4}O_{10}$\\
\hline
$O_9 $&$0   $&$-iO_{10}$&$-iO_{12}$&$0$&$0$&$-iO_{15}$&$-iO_{14}$&$0$&$iO_3$&$0$&$iO_4$&$0$&$iO_8$&$iO_7$&$0$\\
\hline
$O_{10} $&$0   $&$iO_9$&$iO_{11}$&$0$&$0$&$iO_{13}$&$iO_{16}$&$-iO_3$&$0$&$-iO_4$&$0$&$-iO_7$&$0$&$0$&$-iO_8$\\
\hline
$O_{11}$&$ 0 $&$-iO_{12}$&$-\frac{i}{4}O_{10}$&$iO_{16}$&$-iO_{13}$&$0$&$0$&$0$&$iO_4$&$0$&$\frac{i}{4}O_3$&$\frac{i}{4}O_6$&$0$&$0$&$-\frac{i}{4}O_5$\\
\hline
$O_{12}$&$0$&$iO_{11}$&$\frac{i}{4}O_9$&$iO_{14}$&$-iO_{15}$&$0$&$0$&$-iO_4$&$0$&$-\frac{i}{4}O_3$&$0$&$0$&$-\frac{i}{4}O_5$&$\frac{i}{4}O_6$&$0$\\
\hline
$O_{13}$&$-iO_{16}$&$-iO_{15}$&$0$&$0$&$iO_{11}$&$-\frac{i}{4}O_{10}$&$0$&$0$&$iO_7$&$-\frac{i}{4}O_6$&$0$&$0$&$0$&$\frac{i}{4}O_3$&$\frac{i}{4}O_2$\\
\hline
$O_{14}$&$iO_{15}$&$iO_{16}$&$0$&$-iO_{12}$&$0$&$0$&$\frac{i}{4}O_9$&$-iO_8$&$0$&$0$&$\frac{i}{4}O_5$&$0$&$0$&$-\frac
{i}{4}O_2$&$-\frac{i}{4}O_3$\\
\hline
$O_{15}$&$-iO_{14}$&$iO_{13}$&$0$&$0$&$iO_{12}$&$\frac{i}{4}O_9$&$0$&$-iO_7$&$0$&$0$&$-\frac{i}{4}O_6$&$-\frac{i}{4}O_3$&$\frac{i}{4}O_2$&$0$&$0$\\
\hline
$O_{16}$&$iO_{13}$&$-iO_{14}$&$0$&$-iO_{11}$&$0$&$0$&$-\frac{i}{4}O_{10}$&$0$&$iO_8$&$\frac{i}{4}O_5$&$0$&$-\frac{i}{4}O_2$&$\frac{i}{4}O_3$&$0$&$0$\\
\hline
\end{tabular}
\end{center}
\caption{Table of commutators displaying a closed algebra of fifteen operators $O_i$ of the SU(4) group for a pair of qubits. Each entry provides the commutator $[O_i,O_j]$. The seven zeroes in any row or column point to sub-groups SU(2) $\times$ U(1) $\times$ SU(2). Other sets of eight and ten that close under commutation give similarly SU(3) and SO(5) sub-groups, respectively. From \cite{ref7,ref8,ref46}.}
\end{table*}

\begin{table}
\begin{center}
\begin{tabular}{|c|c|c|c|c|c|c|c|c|c|c|c|c|c|c|}

\hline
$O_3 $&$O_{10} $&$O_9 $&$O_2 $&$O_4 $&$O_{12} $&$O_{11}$ & $O_6$&$O_8$&$O_{14}$&$ O_{16}$&$O_5$&$O_7$&$O_{15}$&$O_{13}$\\ 
\hline 
$\frac{1}{2}\tau_z $&$\frac{1}{2}\tau_y $&$\frac{1}{2}\tau_x $&$\frac{1}{2}\sigma_z$&$\frac{1}{4}\tau_z \sigma_z$&$\frac{1}{4}\tau_y\sigma_z $&$\frac{1}{4}\tau_x\sigma_z $&$\frac{1}{2}\sigma_y $&$\frac{1}{4}\tau_z\sigma_y $&$\frac{1}{4}\tau_y \sigma_y$&$\frac{1}{4}\tau_x\sigma_y$&$\frac{1}{2}\sigma_x $&$\frac{1}{4}\tau_z \sigma_x$&$\frac{1}{4}\tau_y\sigma_x $&$\frac{1}{4}\tau_x \sigma_x$\\
\hline
ZI & YI & XI & IZ & ZZ & YZ & XZ & IY & ZY & -YY & XY & IX & ZX & YX & XX \\
\hline
[0100]&[1000]&[1100]&[0001]&[0101]&[1001]&[1101]&[0010]&[0110]&[1010]&[1110]&[0011]&[0111]&[1011]&[1111]\\
\hline
(0101)&(1110)&(1011)&(0010)&(0111)&(1100)&(1001)&(1010)&(1111)&(0100)&(0001)&(1000)&(1101)&(0110)&(0011)\\
\hline
$-\frac{i}{2}\gamma_1\gamma_2$&$-\frac{i}{2}\gamma_3\gamma_1$&$-\frac{i}{2}\gamma_2\gamma_3$&$\frac{1}{2}\gamma_4$&$\frac{i}{4}\gamma_5\gamma_3$&$\frac{i}{4}\gamma_5\gamma_2$&$\frac{i}{4}\gamma_5\gamma_1$&$-\frac{i}{2}\gamma_5\gamma_4$&$\frac{1}{4}\gamma_3$&$\frac{1}{4}\gamma_2$&$\frac{1}{4}\gamma_1$&$-\frac{1}{2}\gamma_5$&$-\frac{i}{4}\gamma_3\gamma_4$&$-\frac{i}{4}\gamma_2\gamma_4$&$-\frac{i}{4}\gamma_1\gamma_4$\\
\hline
$\Sigma_3$ & $\Sigma_2$ & $\Sigma_1$ & $\gamma_4$ & $A_3$ & $A_2$ & $A_1$ & $\alpha_5$ & $\gamma_3$ & $\gamma_2$ & $\gamma_1$ & $\gamma_5$ & $\alpha_3$ & $\alpha_2$ & $\alpha_1$ \\ 
\hline
$G_{03}$ & $G_{02}$ & $G_{01}$ & $G_{30}$ & $G_{33}$ & $G_{32}$ & $G_{31}$ & $G_{20}$ & $G_{23}$ & $G_{22}$ & $G_{21}$ & $G_{10}$ & $G_{13}$ & $G_{12}$ & $G_{11}$ \\ 
\hline
$-i$ & $-Kj$ & -$Kk$ & $i$ & $\pm I$ & $Kk$ & $Kj$ & $Ki$ & $-K$ & $k$ & $j$ & $K$ & $-Ki$ & $-j$ & $-k$ \\
\hline
\end{tabular}
\end{center}
\caption{Dictionary for the fifteen operators $O_i$ of the two-qubit system \cite{ref7,ref8,ref46} in alternative languages: as direct products of individual Pauli matrices of the two spins in the second row and same in shorthand in third row and in allied binary notation with square brackets in the fourth row; as Dirac gamma matrices in the sixth row and in other combinations of Dirac matrices in the seventh \cite{ref12,ref22}, next in bivectors $G_{ij}$ of \cite{ref13}; and in complex quaternions $(i, j, k)$ with $K$ an independent square root of $-1$ in the last row along with an allied binary in round brackets in the fifth row.}
\end{table}

\subsection{The SU(2) $\times$ U(1) $\times$ SU(2) ``Fano sub-group" symmetry and $X$-states}

As mentioned above, an interesting sub-group or sub-algebra of the 15 generator SU(4) is provided by a subset of seven of them and plays a role in many physical systems. To identify them, Table I shows that each of the 15 operators $O_i$ can serve as a non-trivial center U(1) since it commutes with six others. Take as an example the operator $ZZ$ or $O_4$ which we will use as a running example in later sections but emphasize that any of fifteen choices can serve. Its six companions in such a sub-group are $(IZ, ZI, XX, YY, XY, YX)$, that is, $(O_2, O_3, O_{13-16})$. For charged spin-1/2 particles in an external magnetic field, the Hamiltonian in an external magnetic field along the $z$-axis with scalar couplings $(XX, YY, ZZ)$ and what are termed cross-coherences $(XY, YX)$ provides a physical situation with this sub-symmetry \cite{ref7}. It is realized in the CNOT quantum logic gate constructed out of two Josephson junctions \cite{ref47}. Another example is to take as center $ZI$ or $O_3$. Now the other six are $(IX, IY, IZ, ZX, ZY, ZZ)$ or $(O_5, O_6, O_2, O_7, O_8, O_4)$ or, more compactly, $(\vec\sigma, \tau_z \vec\sigma)$. That is, all three Pauli matrices of the first spin along with their multiplication by (any) one of the matrices of the second such as $\tau_z$, clearly provide six that commute with that center $ZI$. These two examples differ, however, in their quantum entanglement properties. 

While they do not as they stand split into two sets of three that mutually commute with each other, the linear combinations $ \frac{1}{2}(I \pm Z) \otimes (X, Y, Z)$, that is, the triplet $\frac{1}{2} (IX+ZX, IY+ZY, IZ+ZZ)$ and similar with minus signs indeed give two such sets of triplets that obey SU(2) commutation relations within themselves while each member commutes with all three of the other set. The operators $ \frac{1}{2}(II \pm ZI)$ behave like projection operators. The very presence of a non-trivial center along with the trivial unit center points immediately to such projection operators and a division of the space into two separate ones, termed generically $P$ and $Q$ with $P+Q=1, P^2=P, Q^2=Q, PQ=QP=0$. For the purposes of unitary integration, since only commutation relations enter, the evolution $U(t)$ does indeed simplify into two independent factors as in Eq.~(\ref{eqn4}) in what may be termed these pseudo-spin SU(2)s \cite{ref7}. The terminology pseudo is invoked because each member of the triplet no longer squares to unity as with Pauli spinors but into the overall commuting objects $ \frac{1}{2}(II \pm ZI)$, involving both centers. Note that in more general contexts beyond our current one of multiple qubits, the very presence of a non-trivial center that squares to unity leads to such a decomposition into orthogonal $P$ and $Q$ spaces. While all fifteen $O_i$ lead to such a separation into two complementary projected spaces within a set of seven generators, only the nine involving both spins can accommodate quantum entanglement as will be discussed further below. Explicitly, when $ZZ$ is the U(1) center, the two mutually commuting SU(2) triplets are $\frac{1}{2} (XX-YY, XY+YX, IZ+ZI)$ and $\frac{1}{2} (XX+YY, XY-YX, IZ-ZI)$. They square to $ \frac{1}{2}(II \pm ZZ)$ and again behave like pseudo-spins but cannot be written in the same factorized form of $\otimes$ of the two spins as at the beginning of this paragraph for the center $ZI$ (or for any choice of center with only one of the spins).           

The above discussion for operators and generators of a sub-group symmetry of SU(4) applies also to the states of a two-qubit system. In a matrix representation, they are also represented by 4 $\times$ 4 Hermitian matrices. The general $4 \times 4$ density matrix of pure or mixed states is characterized by 15 parameters, 3 real ones along the diagonal and 6 complex off-diagonal elements of a Hermitian matrix. It was natural in the original heuristic definition \cite{ref9} to call those with only two non-zero off-diagonal entries, namely those on the anti-diagonal, as $X$-states from visual appearance as in Eq.~(\ref{eqn3}). There are now seven parameters in all and indeed provide an instance of the previous paragraph's SU(2) $\times$ U(1) $\times$ SU(2) sub-group symmetry \cite{ref7}. Depending on the center U(1), the density matrix may or may not look like the letter X but this symmetry perspective shows their commonality \cite{ref10}. And, under operations also by members of that same set of seven operators, the X character is preserved of the physical system. This proves very convenient in many physical applications in reducing the number of parameters and operators to handle. This accounts for the popularity of discussing such $X$-states of a two-qubit system. It also points to a natural extension to higher multiples of qubits and of higher dimensional qudits. We will take this up in Sec. III D. We note that both the Lie algebraic aspect that the seven operators close under commutation and their Clifford algebraic structure that they close under multiplication are important \cite{ref10}. A mathematical description of the occurrence of such sets because of Clifford groups is in \cite{ref48,ref49}. 

Analytical handling is also simplified, reducing to no more than evaluating traces because all the $O_i$ are traceless and square to unity. With any such subset of seven $\{X_i\}$ out of the $O_i$, the density matrix that remains invariant under their operations can be rendered as a linear superposition of them,

\begin{equation}
\rho =(I + \sum_i g_i X_i)/4,
\label{eqn6}
\end{equation}
in analogy to that for a single spin, $(I+\sum_i g_i \sigma_i)/2$. The seven real coefficients $g_i$ in the sum in Eq.~(\ref{eqn6}) parametrize $X$-states and are given by Tr[$\rho X_i$]. Eigenvalues, and entanglement or other correlation properties, can be expressed compactly in terms of them \cite{ref10}. Further, the triplet structure of the Lie-Clifford algebra is most conveniently and geometrically captured by Fig. 5 \cite{ref10,ref46,ref50,ref51}. 

\begin{figure}
\centering
\scalebox{1.6}{\includegraphics[width=2.7in]{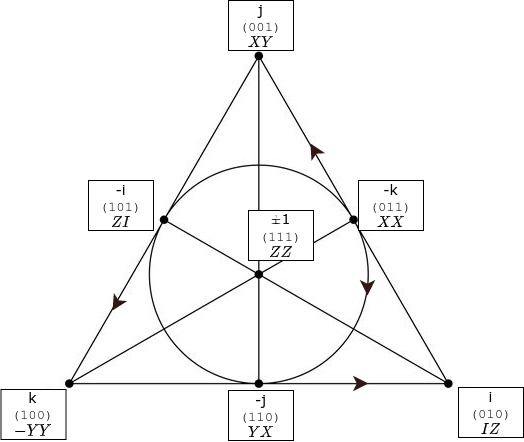}}
\caption{The multiplication diagram for the seven operators that underlie $X$-states. Resembling the Fano Plane, each operator stands on three lines, and each of the seven lines, including the inscribed circle, has on it three operators. All lines are equivalent in finite projective geometry as are all points. On the interior medians, the product of any two operators gives the third, these objects commuting. On the remaining four lines, the operators anticommute, and the product of any two gives cyclically the third with a multiplicative $\pm i$, the plus (minus) depending on the direction of (along/against) the arrow. In the Fano Plane, all seven lines would be arrowed. The points are shown in quaternionic (Sec. III F), binary (Sec. III E), and two-qubit  labelling. With quaternions, the center is $\pm 1$ but for two-qubits any of the fifteen generators can occupy that center, $ZZ$ shown as an illustrative choice. Endpoints of medians are related by a change in sign of the quaternions and a $I \leftrightarrow Z, X \leftrightarrow Y$ ``duality" in the qubit generators. Adapted from \cite{ref10,ref46,ref50,ref51}.}
\end{figure}

This figure, a beautifully symmetric pattern on its own, occurs in projective geometry as the ``Fano Plane" \cite{ref20}, where it is described as the finite projective geometry PG(2,2). Arranging the seven operators at the vertices, mid-points of sides, and in-center of an equilateral triangle, the seven lines shown (including the inscribed circle) each pass through three points, providing the multiplication rule for those $\{X_i\}$. The center U(1) element occupies the in-center of the triangle. On three un-arrowed median lines, all three operators mutually commute, so that the product of two gives the third regardless of order. On the four arrowed lines, the operators mutually anticommute so that the product of two gives $(\pm i)$ times the third, with plus (minus) signs along (against) the sense of the arrow \cite{ref10,ref46}. They may be termed ``cyclic" to contrast with ``commuting" (also called isotropic in \cite{ref52} and orthogonal in \cite{ref53}) for the other set of three medians. The central element commutes, of course, with all six of the others. Each of those six has one `conjugate' element with which it commutes and four with which it anticommutes. All of this can be read off by merely glancing at Fig. 5 which provides simple rules for their manipulation when calculating entanglement and discord \cite{ref10}. Indeed, this figure may be regarded as a direct extension of the ``$(i,j,k)$ cycle" familiar for multiplication or commutation of the Pauli operators for a single qubit. (Also in vector product and quaternion multiplication rules.) It seems natural to call this Fig. 5 the ``Fano triangle" after its Italian geometer originator Gino Fano and the sub-group symmetry of two-qubits as the ``Fano sub-group" in addition to its designation as the Fano Plane of finite projective geometry \cite{ref20}.

Besides the labels of the $X_i$ shown, Fig. 5 also displays a binary and a quaternionic labelling to be discussed in subsequent sub-sections below. For the purpose of this later discussion of quaternionic groups, note the placement of the $ijk$ cyclic triplet on four lines, with one minus sign on the three edges and three minus for the circle, the cyclicity arrow being in opposite senses between them. Extension to octonions that have seven independent square roots of -1 and all seven lines arrowed will be taken up at the end of Sec. III. Also, the finite projective geometry PG( 2, 2) of 7 points and 7 lines with a complete duality between points and lines differs interestingly from the finite Euclidean geometry EG(2, 2) of four points and six lines obtained by dropping the midpoints in the diagram and terminating the median lines at the center. In finite geometries, only the points matter, not the continuous lines connecting them, and while two points define a line in Euclidean, three do so in projective geometries. Another perspective is that the midpoints in Fig. 5 are points ``at infinity" as is the circle a line at infinity in Euclidean terms but projective geometry makes no distinction between points at infinity and ``regular" points at finite location.

In Dirac language, a sub-set of seven operators in a Fano sub-group, as, for example, the one at the beginning of this section with $ZZ = O_4$ the center, are one A, three V, and three T of the other three indices to the one chosen in A. Other possibilities are one V plus three each of T and A, or P plus the six T. Classify the Dirac matrices into five groups: three $\gamma_i$, three $A_i = i\gamma_5 \gamma_i$, three $\alpha_i =-i\gamma_i \gamma_4$, three $\Sigma_i = -i\gamma_j \gamma_k$ (cyclic), and the three singletons  $A_4 = \gamma_4$, $\gamma_5$,  $\alpha_5 = -i\gamma_5 \gamma_4$. In terms of these five classes, the fifteen Fano sub-group sets are $(\gamma_i; \Sigma_i, \alpha_j, \alpha_k, A_j, A_k, \alpha_4)$, $(A_i; \Sigma_i, \gamma_j, \gamma_k, \alpha_j, \alpha_k, \gamma_4), (\alpha_i; \Sigma_i, \gamma_5, \gamma_j, \gamma_k, A_j, A_k)$, and $(\Sigma_i; \gamma_4, \gamma_5, \gamma_i, \alpha_i, A_i, A_4), (\gamma_4; A_{1-3}, \Sigma_{1-3})$, $(\gamma_5; \alpha_{1-3}, \Sigma_{1-3}), (\alpha_5; \gamma_{1-3}, \Sigma_{1-3})$. In each set of seven, the first entry separated by a semi-colon is the commuting U(1) element. A glance at the sets shows involvement of the five classes in natural symmetric patterns. Interestingly, the division of 15 Dirac gamma matrices into five classes, four triplets or vector quantities and three scalar ones, parallels a geometric discussion where twelve are numbered numerically and three with alphabets a, b, and c \cite{ref53} or an analogous division among the generators of the group symmetry of the hydrogen atom \cite{ref21}. These connections between widely disparate problems may be worth further exploration.

While each $O_i$ acting as center gives 15 different $X$-states, they differ in terms of quantum entanglement which rests on cross-correlation between the 1-2 and 3-4 sub-spaces of each spin of the two-qubit system in the canonical basis. When the $O_i$ is a single spin operator in Table I, it does not mix these two spaces and the projection operators provided by such a center do not describe entanglement. As an example, neither $O_2 = IZ$ nor $O_3 = ZI$ diagonal operators, with (1, 1, -1, -1) and (1, -1, 1, -1) entries along the diagonal, respectively, has entanglement whereas $O_4 = ZZ$ with (1, -1, -1, 1) $X$-states may display entanglement for certain values of the parameters in the density matrix. The first of the three diagonal operators acts as unit operators of opposite sign in 1-2 and 3-4 spaces of the two qubits, the second similarly within 1-3 and 2-4 which are the spaces of same spin orientation, up or down. It is the third with center $ZZ$ and grouping 1-4 and 2-3 that pairs a qubit with the other of opposite spin. This simultaneous involvement of both particle and spin seems necessary for quantum entanglement. There is a striking correspondence to Dirac theory where the lower 3 and 4 components of negative energy electron states are reinterpreted as positive energy positron states with a similar spin-flip involved, the 4 seen as up and 3 as down spin of the positron (Sec. 3.10 of \cite{ref22}). Thus, charge conjugation in that context is the analog of entanglement of two qubits.   

Such a sub-division of the 15 into 6 + 9, with single and double spin centers, the former always separable while the latter may admit entanglement, has also been discussed in detail from a finite geometric perspective \cite{ref49}. 15 different Fano planes are listed in their Appendix A and a particular type of geometric hyperplanes called perp-sets identified as a symplectic polar space of rank 2 and order 2, W(3, 2). Depending on a unique quadric $Q_0$ of this space that involves only non-trivial Pauli matrices and whether the perp sets intersect that quadric tangentially or transversally distinguishes the groups of 9 and 6, respectively. An $X$-state set such as $(ZZ, IZ, ZI, XX, YY, XY, YX)$ at the beginning of this section is described in that language as one vector $ZZ$ orthogonal (in place of commuting in Lie algebraic language) to the other six \cite{ref52}. See further discussion in Sec. IV but note the simpler perspective provided in spin/qubit language in terms of the nature of the center, whether a single or double spin operator. In terms of Dirac matrices, it is the nine $(\gamma_i, A_i, \alpha_i)$ as centers that exhibit entanglement not the other six of $\Sigma_i$ and singletons enumerated in a paragraph above.

\subsection{The SO(5) ``Desargues sub-group" symmetry}

Identifying sub-group symmetries other than the Fano sub-group of the previous section proceeds again through the commutator Table I and picking subsets that close under commutation as triplets. As mentioned, such a closed sub-algebra is all that is required for efficient construction of the evolution operator $U(t)$. Thus, $(O_2, O_3, O_{13-16}, O_5, O_6, O_{11}, O_{12})$ is such a set. Atomic and molecular four-level systems often have Hamiltonians that involve only ten parameters  because of dipole selection rules for transitions between the four states.  As a result, two parameters characterize energy positions along the diagonal as in the case of two identical qubits when they share the same energy separation, and four complex off-diagonal dipole couplings display such a sub-group symmetry. Together then, ten real parameters define such a system \cite{ref34}. It is the symmetry SO(5) of five-dimensional rotations. (Actually, it is the double covering group Spin(5) just as SU(2) is such a cover of SO(3), but the distinction is unimportant for most of our discussion.) 

As in the previous SU(2) $\times$ U(1) $\times$ SU(2) Fano sub-group, there are many such SO(5) that can be identified in Table I. Indeed, the above set of ten operators when compared with a similar set in Sec. III A has the first six common while the previous center of that Fano sub-group has been removed and replaced with the last four. This points to a systematic way of picking out the SO(5) examples just as before for SU(2) $\times$ U(1) $\times$ SU(2) . Again, for every $O_i$ in Table I, pick the six other zeroes in that row or column and supplement by four others as required to close the sub-algebra. In terms of Dirac matrices, the above mentioned set of ten are four of the $\gamma$, with indices 1, 2, 4, and 5 and their pairwise combinations. That is, three each of V, A, and T plus P. There is no involvement of the $\gamma_3$. On the other hand, an alternative set of ten $(O_2, O_3, O_{13-16}, O_7, O_8, O_9, O_{10})$ with the same initial six but a different set of four to replace the $O_4$ element is V+T with no involvement of $\gamma_5$ or any pseudoscalar aspect. Yet another example is A+T in the language of Dirac matrices.

A nice geometric object of an equilateral triangle with inscribed circle, with seven line triplets of seven operators (Fig. 5) provided a rendering of the Fano sub-group in Sec. III A. Similarly, the well-known Desargues diagram of projective geometry \cite{ref54,ref55} putting ten points on ten lines gives a rendering of the SO(5) sub-group which may, therefore, be called the ``Desargues sub-group." Various renderings are in \cite{ref46,ref50} and in \cite{ref56} whose Fig. 5 refers to it as the ``Petersen" graph, dual to a five-point ``ovoid," these objects to be discussed further below in Sec. IV. Yet another geometric alternative that follows the previous paragraph's prescription of dropping the center in the set of seven and adding four others is to remove the in-center and add a new vertex off the plane of the triangle, along with its edges to the other three already extant vertices and corresponding three mid-points. This gives the next order simplex to the 2-simplex triangle, namely the 3-simplex tetrahedron, to represent the SO(5) Desargues sub-group. The ten lines are the six edges and four face circles of a tetrahedron to be shown and discussed below in Sec. III C.

Turning now to the evolution $U(t)$, as per Eq.~(\ref{eqn4}), for such a SO(5), the Hamiltonian with $N=4$ is most naturally chosen as $n= N-n =2$, so that all handling is of 2 $\times$ 2 block matrices. For the ten-parameter $H$, a convenient representation \cite{ref34} is $H(t)=F_{21}\sigma^{(2)}_z-F_{31}\sigma^{(2)}_y+F_{32}\sigma^{(2)}_x-F_{4i}\sigma^{(1)}_z\sigma^{(2)}_i+F_{5i}\sigma^{(1)}_x\sigma^{(2)}_i-F_{54}\sigma^{(1)}_y$, where the ten arbitrarily time-dependent coefficients $F_{\mu \nu}(t)$ form a $5 \times 5$ antisymmetric real matrix in keeping with the aspect of five-dimensional rotations. (We will use $\mu,\nu=1-5$ and $i,j,k=1-3$ and summation over repeated indices.) As noted, several quantum optics and multiphoton problems of four levels driven by time-dependent electric fields have such a Hamiltonian. It has also been considered extensively in coherent population transfer in many molecular and solid state systems \cite{ref57}. Casting this Hamiltonian in the form of Eq.~(\ref{eqn5}), we have

\begin{equation}
{\bf H}^{(1,2)}=(\mp F_{4k} -\frac{1}{2} \epsilon_{ijk} F_{ij}) \sigma_k, {\bf V}=iF_{54} {\bf I}^{(2)}+F_{5i} \sigma_i.
\label{eqn7}
\end{equation}

The 2 $\times$ 2 block matrix $\bf{z}$ in Fig. 4 obeys a matrix Riccati equation and the four entries ($\mu = 1-4$) can be chosen as real: $ z_{\mu}=z_4,z_i$: ${\bf z}=z_4 {\bf I}^{(2)} -iz_i \sigma_i$. The equation for $z_{\mu}$ takes the form \cite{ref34} 

\begin{equation}
d{z}_{\mu}/dt = F_{5\mu}(1-z_{\nu}^2)+2F_{\mu \nu}z_{\nu}+2F_{5\nu}z_{\nu}z_{\mu}.
\label{eqn8}
\end{equation}
(As an alternative, ${\bf V}$ and $ {\bf z}$ can also be rendered in terms of quaternions $(1,-i\sigma_i)$). We can now construct a five-dimensional unit vector $\vec{m}$ out of the four real $z$,

\begin{equation}
m_{\mu}=\frac{-2z_{\mu}}{(1+z_{\nu}^2)}, \,m_5=\frac{(1-z_{\nu}^2)}{(1+z_{\nu}^2)},\,\,\,\, \mu,\nu=1-4. 
\label{eqn9}
\end{equation}
The nonlinear Eq.~(\ref{eqn8}) in $z$, becomes of simple, linear Bloch-like form, 

\begin{equation}
d{m}_{\mu}/dt = 2F_{\mu\nu}m_{\nu},\,\,\,\,\, \mu,\nu=1-5,
\label{eqn10}
\end{equation}
which is the obvious analog of the Bloch equation of a single spin involving the cross product, now in its higher dimensional antisymmetric counterpart for rotations in five dimensions. 

\begin{figure}
\centering
\vspace{-1in}
\scalebox{1.5}{\includegraphics[width=3.5in]{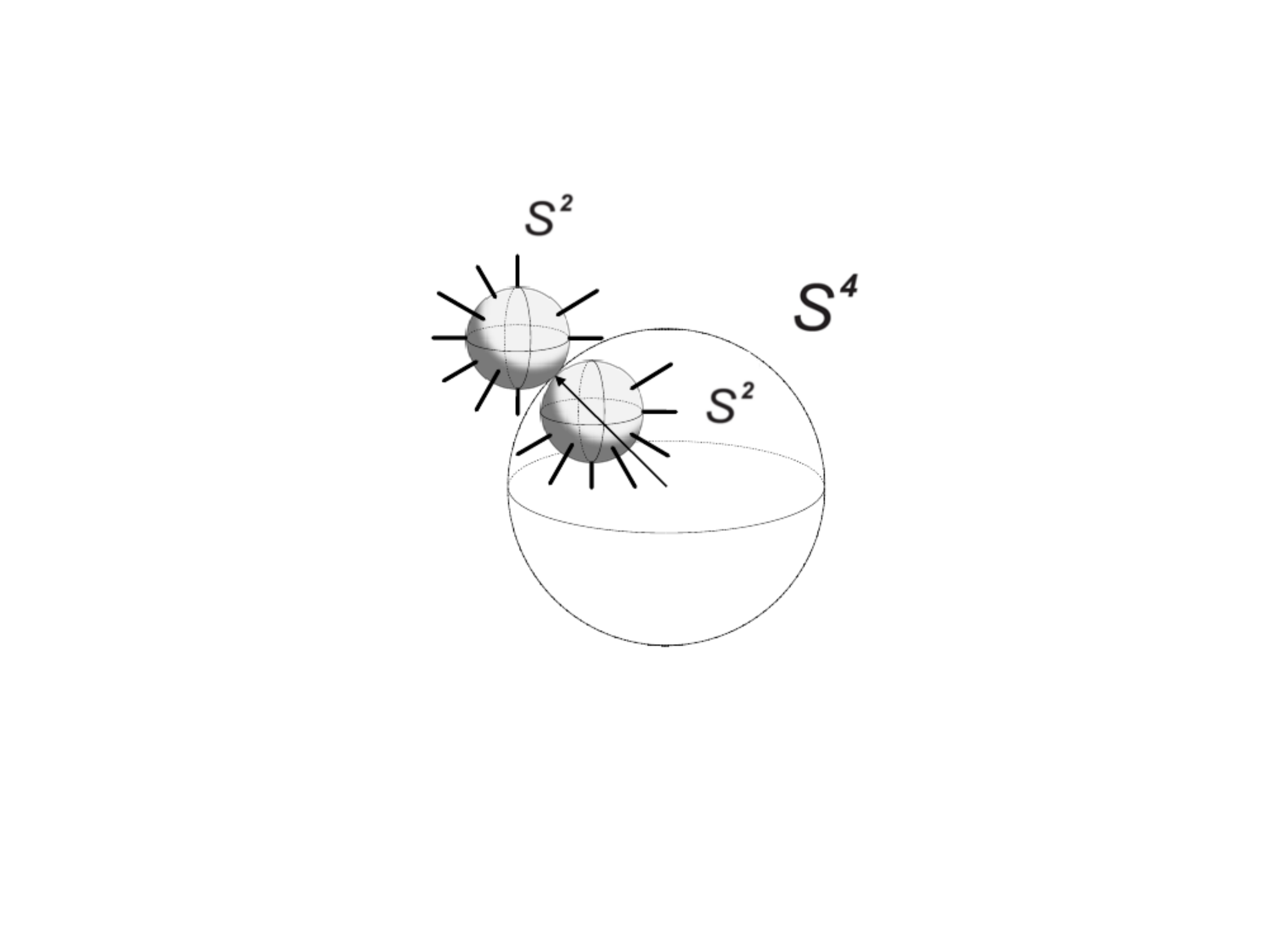}}
\vspace{-1in}
\caption{Analogous to Fig. 2, the fiber bundle for a two-qubit system that involves an so(5) sub-algebra of the full su(4). The base manifold now is a four-sphere $S^4$, at each point of which is a six-dimensional fiber consisting of two spiked-spheres of su(2) as in Fig. 2 \cite{ref34}.}
\end{figure}

Solving this provides the first two factors in Fig. 4 and also the effective Hamiltonian for the two diagonal blocks of the third factor which may in turn be analyzed as a spiked Bloch sphere of a single SU(2) \cite{ref34}.
As in the single-spin case, one can do an inverse stereographic projection, now from the four-dimensional plane $z \in R^4$ to the four-sphere $S^4$. It provides a higher-dimensional polarization vector for describing such two spin problems. In all, such Hamiltonians possessing Spin(5) symmetry are, therefore, described by the geometrical picture of one S$^4$ and two S$^2$ spheres along with two phases, as shown in Fig. 6 and the unitary evolution operator depicted as in Fig. 4. The former base manifold is similar albeit of higher dimension than that of a single qubit while the fiber now is a six-dimensional object and not a single phase. It can be nicely pictured as two spiked Bloch spheres sitting on each point of the base manifold. Although a much larger, ten-dimensional, object than in the single spin case, it is nevertheless an elegant and easily accessible generalization of Fig. 2 \cite{ref34}.

\subsection{SU(3) sub-groups and the complete SU(4) Hamiltonian involving all fifteen operators}

Other sub-group symmetries of SU(4) include SU(3) with eight parameters that can be thought of as two independent energy parameters along the Hamiltonian's diagonal and three complex off-diagonal couplings. A general three-level system, embedded into four with the fourth level completely uncoupled, constitutes a trivial example of such an su(3) sub-algebra but less trivial examples can also occur. The ${\bf z}$ now has two non-zero complex $z$ for a total of four parameters. The description \cite{ref45} of this four-dimensional manifold, as well as the remaining SU(2) and a U(1) phase, parallel the discussion of the general SU(4) that we now take up. Their manifolds are, however, more complex than spheres and one-dimensional fibers. 

\begin{figure}
\centering
\scalebox{2.0}{\includegraphics[width=1.5in]{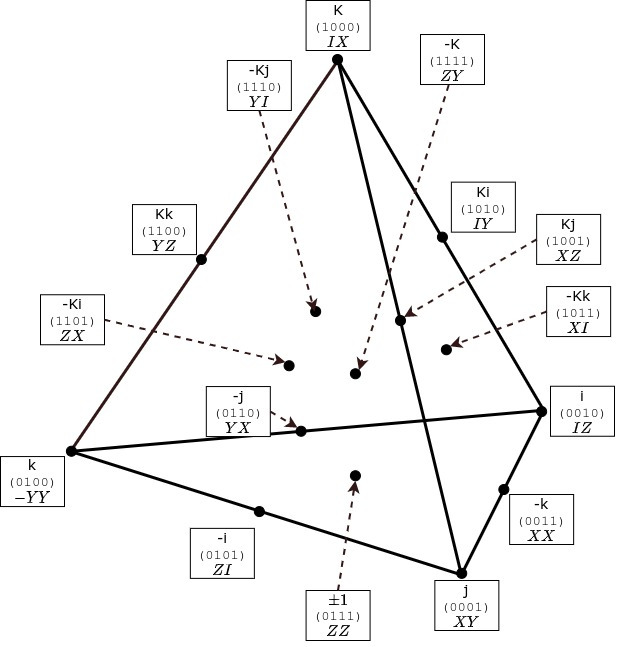}}
\caption{The fifteen generators of two-qubits placed on a tetrahedron (3-simplex) in bi-quaternion, 4-binary, and qubit language. Vertices are the quaternion $(kij)$ with corresponding binary $(0xyz)$ of one non-zero unit entry for the base as in Fig. 5 and an independent imaginary unit $K$ placed at (1000). Midpoints of the six edges and four face centers obtained by binary addition or quaternionic multiplication are shown, with the face center of the base kept as $\pm 1$ as in Fig. 5. The body center is $-K$. Triplet lines, 35 in number, are the 6 edges, 12 medians on the faces, 4 face circles, 4 altitudes, 3 of body center with a pair of opposite midpoints such as $(-i, -K, Ki)$, and 6 that link two face centers with a midpoint such as $(-Kj, -Kk, -i)$. With the choice of qubit-qubit $ZZ$ as center as in Fig. 5, points are also labelled in terms of those generators as per Table II. Dropping the face and body centers gives the 10 point, 10 triplet lines (6 edges and 4 face circles) of the Desargues sub-group of Sec. III B. Note a duality between face and edge midpoints or corners and midpoints, marked by a change in sign of the quaternions or the interchange of generators $I \leftrightarrow Z, X \leftrightarrow Y$. Adapted from \cite{ref46}.}
\end{figure}               

Moving beyond sub-groups to the full SU(4), consider an arbitrary $4 \times 4$ Hamiltonian with its entire complement of 15 operators/matrices. Such a $H$ is obtained by adding to the previous Spin(5) Hamiltonian in Sec. III B five additional terms: $F_{65} \sigma ^{(1)}_z +F_{64} \sigma ^{(1)}_x +F_{6i} \sigma ^{(1)}_y \sigma ^{(2)}_i$. This corresponds to the energy levels being arbitrarily positioned as they would be in a general four-level (not just two-qubit) system, and the two other couplings restored. Correspondingly, Eq.~(\ref{eqn7}) gets an additional term $\pm F_{65} {\bf I}^{(2)}$ in the diagonal ${\bf H}^{(1,2)}$ while in ${\bf V}$, the $F_{5 \mu}$ are replaced by $F_{5 \mu} -iF_{6 \mu}$. Thus, the full SU(4) amounts to a simple modification of the previously considered Spin(5) by adding a term proportional to the unit operator to the diagonal blocks and making the four $F_{5 \mu}$ complex, with $F_{6 \mu}$ absorbed as their imaginary parts \cite{ref34}. A full 6 $\times$ 6 antisymmetric collection of generators (see Appendix B of \cite{ref34}) may then be viewed as the symmetry SO(6) to be discussed further below.
 
With the tetrahedron introduced above for the ten-parameter SO(5), the full fifteen-parameter SU(4) completes that figure by adding also the four face centers and the body center. In turn, the triplet lines that have Lie-Clifford algebra are now 35 in number. Besides the previous ten of edges and face circles, there now are 12 medians, 4 altitudes, 3 lines that link the body center to two midpoints, and 6 that link two face centers to a midpoint. The full tetrahedron is shown in Fig. 7 \cite{ref46}. A similar description is in \cite{ref58}. It is difficult to display some of the 35 lines but Fig. 8 of \cite{ref53} makes a good attempt. Each of the four faces of the tetrahedron is now a Fano subgroup. A less immediately visual one is formed by the six midpoints and body centre of the tetrahedron, the seven lines now being the facial circles and the three connecting opposite midpoints to the body centre. The various Desargues sub-groups of ten points and ten triplet lines (6 edges and 4 face circles, as noted) with their labelling, and the ovoid consisting of complementary five points are shown in \cite{ref58}.

In constructing $U(t)$ now for the full SU(4), the Riccati Eq.~(\ref{eqn8}), now for complex $z$, becomes

\begin{eqnarray}
d{z}_{\mu}/dt & = & F_{5\mu}(1-z_{\nu}^2) -iF_{6 \mu}(1+z_{\nu}^2)+2F_{\mu \nu}z_{\nu} \nonumber \\
 \!\!& + &\!\! 2(F_{5\nu}\!+\! iF_{6\nu})z_{\nu}z_{\mu} \!- \! 2iF_{65}z_{\mu},  \, \, \mu,\nu=1-4.
\label{eqn11}
\end{eqnarray}
 
Just as the very structure of Eq.~(\ref{eqn8}) suggests that $z_{\mu}$ and $(1-z_{\nu}^2)$ with suitable normalization define a five-dimensional unit vector $\vec{m}$ in Eq.~(\ref{eqn9}), so too now for a set of six complex quantities $\vec{m}$. And the nonlinear Riccati equation for the four complex $z_{\mu}$ in Eq.~(\ref{eqn11}) becomes a linear Bloch-like equation as before, now in six dimensions,

\begin{equation}
d{m}_{\mu}/dt = 2F_{\mu\nu}m_{\nu}, \,\,\,\,\, \mu,\nu=1-6.
\label{eqn12}
\end{equation}
Once again, the $m_{\mu}$ obey a first-order equation with an antisymmetric matrix which now describes rotations in six dimensions. Since the 15 $F_{\mu\nu}$ are real, the real and imaginary parts of the six $m_{\mu}$ each obey such a six-dimensional rotational transformation. The geometrical picture now is of a Grassmannian manifold \cite{ref34} as follows. The six complex $m_{\mu}$ obey three constraints and thus describe a nine-dimensional Stiefel manifold St(6, 2, R) with SU(4)/[SU(2) $\times$ SU(2)] symmetry. It differs in a phase parameter from an eight-dimensional Grassmannian manifold Gr(4, 2, C) according to St(6, 2, R) $\simeq$ Gr(4, 2, C) $\times$ U(1). Such a Gr manifold describes the four complex $z_{\mu}$. An alternative view in terms of a five-sphere S$^5$ and two orthogonal six-dimensional unit vectors from the origin rotating within that sphere is given in \cite{ref34}. Yet another description is to use what are called Pl\"{u}cker coordinates, six complex parameters which identify complex hyperplanes of Gr(4, 2, C), again discussed in \cite{ref34}. They have been further discussed for pure states of three qubits, also a system with 15 real parameters, and used \cite{ref59} for transforming between so-called W and GHZ (Greenberger-Horne-Zeilinger) states that are familiar in quantum information \cite{ref59a}. Such a transformation has also been discussed using a SU(2) $\times$ SU(2) sub-group symmetry \cite{ref59b}.   

The occurrence of five- and six-dimensional antisymmetric equations in Eq.~(\ref{eqn10}) and Eq.~(\ref{eqn12}) that are simple generalizations of the familiar vector Bloch equation for a single qubit reflect the isomorphism between the groups SU(4) and SO(6) (more accurately, its covering group Spin(6): SU(4): SO(6) $\sim$ SU(4)$/Z_2$.). They suggest a mapping between the generators of these groups of 15 generators as given in Eq.(B1) of \cite{ref34}. That mapping also extends to the full set of operators that describe the non-relativistic hydrogen atom and its transitions in quantum mechanics \cite{ref21}. Interestingly, this correspondence between SU and SO symmetries is only true of the single- and two-qubit problems, does not hold for any higher number of them. It rests on a curiosity in number theory called the Ramanujan-Nagel theorem, that the Diophantine equation relating squares of integers and integer powers of 2, $2^n=k^2+7$, has solutions in integer $n$ and $k$ for only five values \cite{ref60}.

\subsection{Larger number of qubits and their $X$-states}
 
Recognizing the symmetry group and structure behind $X$-states also permits ready generalization to a larger number of qubits. Indeed, for this purpose, stepping back from two-qubits to a single qubit, any 2 $\times$ 2 density matrix is necessarily of X character! It has, of course, SU(2)  symmetry. The Fano sub-group symmetry of two-qubit $X$-states, SU(2) $\times$ U(1) $\times$ SU(2), may be regarded as repeating the previous one-qubit SU(2) and attaching the center U(1) in between. This view also fits into a 4 $\times$ 4 density matrix as two 2 $\times$ 2 ones of 1-4 and 2-3 spaces in the canonical basis with a mutual phase between the spaces. The specific breakdown into 1-4 and 2-3 as in Eq.~(\ref{eqn3}) and the example in Fig. 5 as against other 2 $\times$ 2 breakdowns will be discussed further below. The generalization is immediate to a higher number, say $q$ of qubits. At each step, two copies of the previous with an added U(1) in between gives the corresponding symmetry and set of $X$-states. Thus, for a system of three qubits, the SU(2) $\times$ U(1) $\times$ SU(2) $\times$ U(1) $\times$ SU(2) $\times$ U(1) $\times$ SU(2) group of 15 generators, a sub-group of the full SU(2$^3$ =8) of 63 generators, is the sub-group symmetry of such three-qubit $X$-states. This corresponds to 7 real diagonal and 4 complex anti-diagonal elements in a $8 \times 8$ matrix. For any $q$, the full symmetry group is SU($2^q)$ with an explosively large number $2^{2q} -1$ of generators but the smaller $2^{q+1} -1$ set provides the $X$-states and their operators. These $q$-qubit $X$-states constitute the finite projective geometry PG($q$, 2) and, for $q=3$, can be geometrically represented by the same 3-simplex tetrahedron in Fig. 7 that was used for all two-qubit generators. Compared to general mixed states, pure states also form a subset with fewer parameters, that number coinciding with $X$-state values above so that again the same figures can be used to represent them.

That $X$-states of two qubits embrace seven whereas of three qubits have 15, which is also the number of generators or independent parameters of a full two-qubit system, links to an interesting nesting of projective geometries. PG(2, 2) of seven sits within PG(3, 2) of fifteen with an additional eight members that may be seen as the vertices of a cube (Fig. 8). This triangle plus cube provides an alternative to the tetrahedron in Fig. 7 to represent all fifteen operators and 35 triplet Lie-Clifford lines of the full SU(4) discussed in Sec. III C. The Fano triangle's seven vertices and seven triplet lines are supplemented by the eight vertices and 28 triplet lines (12 edges, 12 medians of faces, 4 body diagonals) of what is known as the Clifford cube shown in Fig. 8. Such a cube has been used by computer scientists to represent a three-color (RGB) imaging scheme \cite{ref61}. Generalizing to higher $q$, the PG($q$, 2) of $X$-states has $2^{q+1} -1$ points running through the sequence $3, 7, 15, 31, 63, \ldots $, each the previous number of line, triangle, tetrahedron, etc., supplemented by the hypercube of $(q+1)$ dimensions with $2^{q+1}$ vertices: 4, 8, 16, 32, $\ldots$. On the other hand, the total number of parameters for a density matrix of $q$-qubits is the sequence $3, 15, 63, 255, \ldots$ of PG(2$q$-1,2) which is the number of all generators of SU($2^q$) \cite{ref62,ref63}. A binary labelling in the next sub-section provides another convenient addition to these algebraic and geometric views of such sequences. See Eq.~(\ref{eqn13}). The identification of Clifford algebras with PG($n$, 2) over GF(2), the Galois field of order 2, for $n=2-4$ also figured in the mathematics literature from a different approach in \cite{ref64,ref65}. And, independently, by yet another group \cite{ref65a,ref65b} who also made the connection of these projective geometries to qubits and qudits.   

\begin{figure}
\centering
\scalebox{1.5}{\includegraphics[width=3.5in]{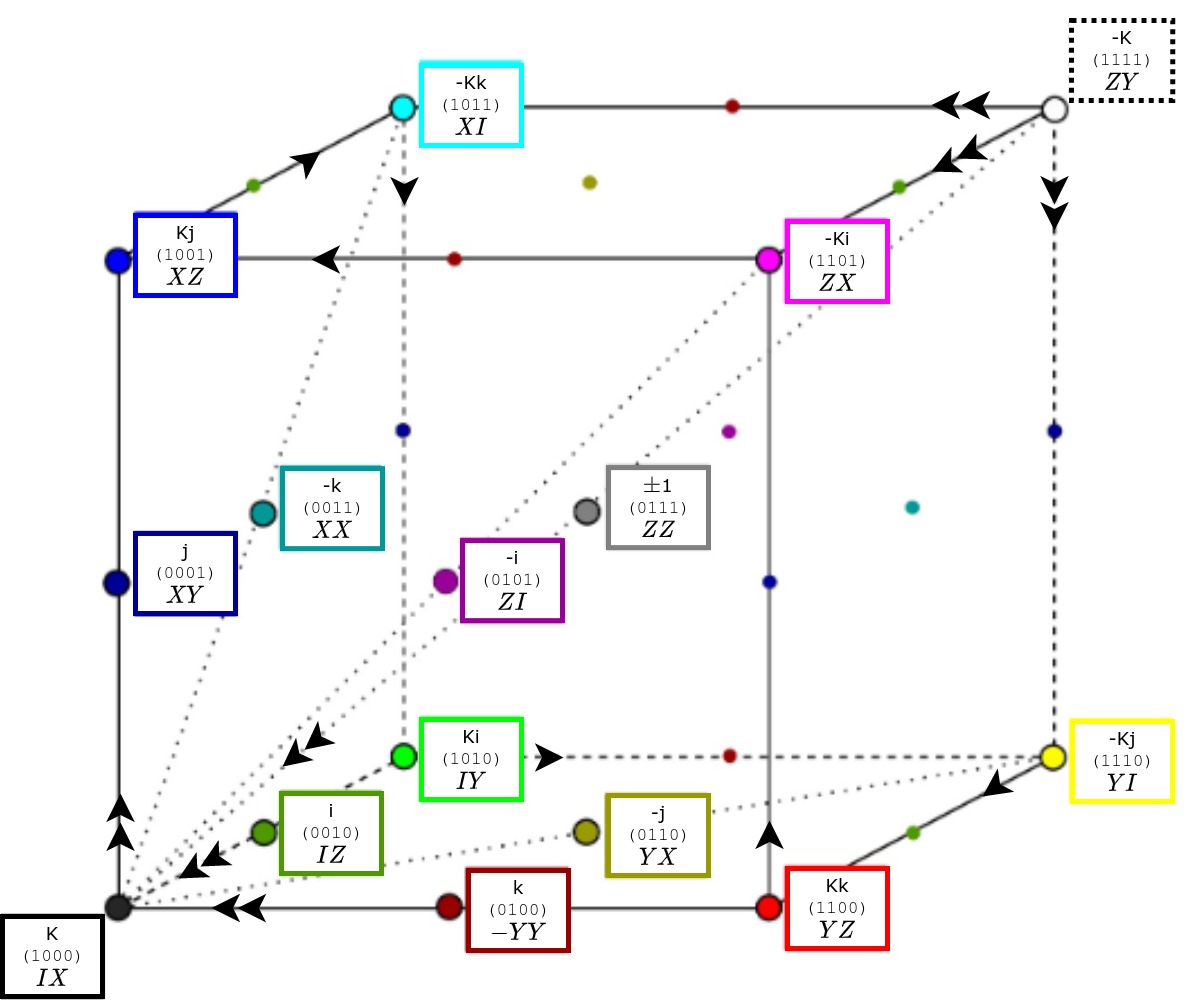}}
\caption{The Clifford Cube as a complement to the Fano Plane/Triangle of Fig. 5. The $x$, $y$, and $z$ axes are laid out along the horizontal, vertical, and into the page directions, respectively. The 3-binary of Fig. 5 is extended to the 4-binary $(txyz)$ by adding the 8 vertices/corners of the cube, starting with (1000) and corresponding new imaginary unit $K$, and placing the $(0xyz)$ as midpoints of edges, faces and the body of the cube to complete it by binary addition or multiplication of quaternions by $K$. Corresponding qubit generators are shown, as also in Table II, for the same choice made before of $ZZ$ as center. Single arrows show a quaternionic multiplicative flow of a circuit connecting six of the vertices, leaving the corners $\pm K$ unvisited. Double arrows at those corners show the sense of multiplication of qubit operators for four ``cyclic" lines there, the other three being commutative, unarrowed lines. Dual $\pm (ijk)$ of opposite sign occur pairwise on midpoints of an edge and its orthogonal face, while the same multiplied by $K$ as opposite corners of the cube. Points left unlabelled to avoid clutter carry the same labels as their corresponding geometric counterparts shown with labels. The 12 triplet lines of the edges and face diagonals, together with 4 body diagonals for a total of 28, supplement the 7 in Fig. 5 to give the full set of 35 for the qubit-qubit pair described in Fig. 7. Adapted from \cite{ref46}.}   
\end{figure}

\subsection{A binary labelling for multiple qubits}

The four generators of unity and the Pauli $\vec{\sigma}$ of a single qubit have a natural 2-binary labelling that is widely used in quantum information: $I : 00, \sigma_z : 01, \sigma_y : 10, \sigma_x : 11$ \cite{ref1}. Extension to multiple qubits is immediate by adding another such pair for each new qubit. Thus $\tau_x \sigma_x$ or, alternatively, $\sigma^{(2)}_x \sigma^{(1)}_x$ or $XX$, is assigned [1111], and a three-qubit register [000010] would represent $I^{(3)} I^{(2)} \sigma_y ^{(1)}$ while [110110] would be the operator $\sigma_x ^{(3)} \sigma_z ^{(2)} \sigma_y ^{(1)}$. Corresponding base-10 values of these binary strings, ranging from 1 to $2^{2q}-1$ and the sequence $3, 15, 63, 255, \ldots$ noted in the last sub-section, would uniquely label states or operators of $q$-qubits \cite{ref46,ref58}. An alternative extension is more economical for the smaller number of $X$-states (also the number for pure states) noted in the previous paragraph. Draw from the above four labels for a single qubit a rule that an initial 0 reads the subsequent entries as $0: I, 1: Z$ whereas an initial 1 reads instead $0: Y, 1: X$ in what follows. In \cite{ref62}, they were named $D_i$ and $A_i$, respectively. This interpretation also contains a natural ``duality" noted throughout this paper of $I \leftrightarrow Z, X \leftrightarrow Y$. 

Extending that rule to larger strings, the seven qubit-qubit $X$-states are rendered as 3-binary strings: $000: I, 001: I Z, 010: Z I, 011: ZZ, 100: YY, 101: YX, 110: XY, 111: XX$. Generalizing to multi $q$-qubits, each step introduces an additional slot in the binary string, grouping all $Z$ operators under $D_i$ and all $(X, Y)$ under $A_i$; for example, the 3-qubit $X$-states have $(Z_1, Z_2, Z_3, Z_1Z_2, Z_2Z_3, Z_3Z_1)$ in the first and $(X_1X_2X_3, Y_1X_2X_3, X_1Y_2X_3, Y_1Y_2X_3, X_1X_2Y_3, Y_1X_2Y_3, X_1Y_2Y_3, Y_1Y_2Y_3)$ in the second for the set of 15 operators involved \cite{ref62}. A variant that is aesthetically a better fit to geometric pictures and to a quaternionic rendering was presented in \cite{ref46} and will be discussed below in Sec. III F.

This more economical $(q+1)$-binary running from 1 to $2^{q+1}-1$, which applies both to $X$-states of $q$-qubits and to pure states of $(q+1)$-qubits, has a natural connection to geometrical pictures and the finite projective geometry PG($q$, 2) discussed in the previous sub-section. Each successive step in $q$ introduces an initial 0 before the previous string along with a new set with an initial 1. Thus, indeed, the sequence 3, 7, 15, $\ldots$ is the previous number plus the number of vertices of a hypercube. Geometrically, to each previous line, triangle, tetrahedron, etc., a new vertex is added in a new dimension represented by the initial 1 which is connected to all the previous vertices. The new vertex and its introduction of mid-points of the connections to the ones of the previous $q$ also matches the prescription when referring to qubit generators given in the previous sub-section of duplicating the generators and adding a single U(1) to get to the next step. The iteration can also be compactly rendered as
 
\begin{equation}
PG(q, 2) = PG(q-1, 2) + EG(q, 2), 
\label{eqn13}
\end{equation}
and its obvious iteration that PG($q$, 2) = EG($q$, 2) + EG($q-1$, 2) + $\ldots$, a string of hypercubes, in conformity with a more general expression for PG($n, m$),

\begin{equation}
PG(n, m) = \frac{m^n-1}{m-1} = m^n + m^{n-1} + \ldots = EG(n, m) +EG(n-1, m) + \ldots .
\label{eqn14}
\end{equation}

\subsection{A quaternionic correspondence}
Hamilton's quaternions, a four-dimensional division algebra, has long been regarded as an alternative to Pauli matrices for describing a quantum spin-1/2 \cite{ref15}. Similar correspondences for multiple qubits also bring out group-theoretic links between the discrete/finite groups of multiple quaternions and the continuous SU($2^q$) groups and generators of the qubit systems. As noted in the Introduction, quaternions are natural for geometric algebra and Klein himself gives a nice description of three-dimensional rotations in terms of them \cite{ref66}. Maxwell too had advocated their use although he himself wrote out his equations for electromagnetism in component form. However, subsequent developments in physics went in a different direction \cite{ref50,ref55}. Vectors, and scalar and cross products of them, became standard whereas geometric algebra and quaternions would not have so separated a single product. This would have had the merit of permitting division as well which is not defined for two vectors in arbitrary directions. Many advantages would have accrued \cite{ref14,ref66}. Interestingly, the consideration of SU symmetries in qubit systems as discussed in this review brought connections naturally \cite{ref50} to finite projective geometries and geometric algebra, along with correspondences between continuous Lie groups and discrete finite groups as we will now discuss. And, in Sec. IV, we will consider the development from the purely geometric approach not motivated by quantum information, but now coming together onto a common platform.

A variant of the more economical binary labelling at the end of Sec. III E but which fits better the geometric diagrams of simplexes and a consistent build up of their labels is the one adopted for the Fano Plane in Fig. 5, the Clifford cube (Fig. 8), and the tetrahedron (Fig. 7). Consistently using round brackets for this binary to distinguish from the earlier one with square brackets, start with the basic triplet of quaternions $(i, j, k)$ or the Pauli matrices $(X, Y, Z)$. With the correspondence $(i, j, k) \rightarrow -i(\sigma_x, \sigma_y, \sigma_z)$ of the two triplets that obey the same cyclic multiplication rules, the 1-simplex of a single qubit or quaternion is a line of three points. With all seven lines of Fig. 5 completely equivalent, including the inscribed circle, any of them can be the starting one. Choose the right edge, labelling the three points with a 2-binary of (01), (10) and (11) as shown. Again, for convenience of generalization, place $-k$ as the mid point, in assonance with the spin language convention that puts $Z$ as the diagonal object in quantum physics. The minus sign is again for later consistent generalization of every further simplex obtained by introducing a step into a new dimension with a new $k$-like square root of minus 1 and its partner $(-k)$-like generalization of mid-point of a line to facial, space/body, etc., center. They will always carry a string of 1's in the round bracket binary label. For these reasons, the correspondence to quaternions is $(xyz) \leftrightarrow (kij)$.

\begin{table}
\begin{center}
\begin{tabular}{|c|c|c|c||c|c|c|c|}

\hline
$1$&$k$&$-1$&$-k$&$i$&$j$&$-i$ & $-j$ \\ 
\hline 
$k$&$-1$&$-k$&$1$&$j$&$-i$&$-j$&$i$ \\
\hline
$-1$&$-k$&$1$&$k$&$-i$&$-j$&$i$&$ j$ \\
\hline
$-k$&$1$&$k$&$-1$&$-j$&$i$&$j$&$-i$ \\
\hline
\hline
$i$&$-j$&$-i$&$j$&$-1$&$k$&$1$&$-k$ \\
\hline
$j$&$i$&$-j$&$-i$&$-k$&$-1$&$k$&$1$  \\
\hline
 $-i$&$j$&$i$&$-j$&$1$&$-k$&$-1$&$k$ \\ 
\hline
 $-j$&$-i$&$j$&$i$&$k$&$1$&$-k$&$-1$ \\ 
\hline
\end{tabular}
\end{center}
\caption{Cayley multiplication table for quaternion group Q$_8$ that can be characterized by two parameters: $a=k, b=i, a^2=b^2=-1$, with $ab=-ba =j$. Note the natural 2 $\times$ 2 block matrix structure with one of the C$_4$ sub-groups in the diagonal blocks. With the mapping $(i, j, k) \rightarrow ( -i\sigma_x, -i\sigma_y, -i\sigma_z)$, the two triplets share the same multiplication rules and the same structure applies to the SU(2) generators.}
\end{table}

Next, the basic quaternion group Q$_8$ \cite{ref67} of the set $\pm (1, i, j, k)$ with its Cayley table shown in Table III can be set in correspondence with the Fano sub-group of seven operators and both placed on the Fano Plane's equilateral triangle as shown in Fig. 5. This step to the 2-simplex introduces a new vertex, which may be denoted $k$, and connected to the previous three points. It joins with the previous endpoints $(i, j)$ to provide the vertices of the Fano triangle. The center of the group, $\pm 1$, is the geometric center and two new mid-points, ``conjugate" negatives of the previous vertices $(i, j)$, arise at this step. Note four cyclic lines (three edges and circle) shown arrowed, and three commuting median lines. As stated before, all lines are equivalent in a finite projective geometry. The circulation of the arrows is counter-clockwise around the edges and clockwise in the inscribed circle. Correspondingly, the extension from 2- to 3-binary labelling proceeds by adding an initial 0 to the points of the 1-simplex, and calling the new vertex (100). The other points then acquire labels by binary addition with the commuting center as (111). In this manner, the Cayley table of Q$_8$ is placed on the Fano Plane. 

Such a 3-binary $(xyz)$ with round brackets labels the points in Fig. 5 on purely geometric grounds, $x=0$ as the right edge, $y=0$ as the left edge, and $z=0$ the base edge. The ascribing of quaternion $(i, j, k)$ to the points is to some extent arbitrary since all points and lines are equivalent, and related by simple geometric transformations such as rotations in the plane. However, it is natural to place $\pm 1$ and, equivalently (000)/(111), as the center of the triangle. Pairs of opposite signs then stand on opposite ends of the medians, conjugates under binary addition. As stated, it proves convenient also for what follows to standardize the new vertex as $k$-like when proceeding to generalization to higher dimensions or number of qubits. Another perspective provided by geometric and Clifford algebra is that 1 is a scalar, $(i, j, k)$ a vector, while $(-i, -j, -k)$, formed out of antisymmetric pairwise products, is a bi-vector, and $ijk=-1$ is a pseudoscalar. Together, they are placed on the Fano Plane in Fig. 5.

With the equivalence of multiplication rules between quaternions and Pauli matrix generators of SU(2), a similar placement can be made of 1-qubit generators with $(I, -i\sigma_x, -i\sigma_y, -i\sigma_z)$ and $(i\sigma_x, i\sigma_y, i\sigma_z, -I)$ labelling points in Fig. 5. Such an assignment also occurs as Fig. 1 of \cite{ref65a}. However, a sign is irrelevant when dealing with generators of a continuous group. Instead, 2-qubit generators of the Fano sub-group provide such a correspondence and a natural identification with the 7-generator Fano sub-group and $X$-states of Sec. III A. Again, since any $O_i$ can serve as center, the example shown in Fig. 5 is for $ZZ$ as center, with the remaining six as commuting pairs at opposite ends of the medians. However, this match to 2-qubit generators is not to Q$_8$ as such but to another closely related order-8 ``co-quaternion" group that is isomorphic to the dihedral group D$_4$. This will be discussed further at the end of this sub-section. Note that the change in sign of a quaternion has as its counterpart the duality exchange $I \leftrightarrow Z, X \leftrightarrow Y$. 

The same quaternion and spin generator labelling in Fig. 5 is shown in Table II, and it must again be noted that it is for the specific example chosen, other centers and choices of seven generators yielding other correspondences between quaternions and generators. As noted, the example chosen corresponds to a physical set up of two spins in a magnetic field along the $z$-axis along with the four operators of magnetic interactions in the orthogonal $X-Y$ plane. Because any $O_i$ can serve as center and placed as the geometric center, their square bracket and double spinor names cannot be universally related 1:1 to the geometric round bracket labels \cite{ref46}. Also, both sets of 4-binary can be rendered in base ten to run as a single number so that points in Figs. 7 and 8 may be labelled from 1 to 15 as in \cite{ref58} and the 2-qubit generators as in \cite{ref46} although we have retained the $O_i$ names for them because of previous usage in \cite{ref7,ref8,ref46}.

\begin{table}
\begin{center}
\begin{tabular}{|c|c|c|c||c|c|c|c|}

\hline
$1$&$k$&$-1$&$-k$&$Ki$&$Kj$&$-Ki$&$-Kj$ \\ 
\hline 
$k$&$-1$&$-k$&$1$&$-Kj$&$-Ki$&$Kj$&$Ki$ \\
\hline
$-1$&$-k$&$1$&$k$&$-Ki$&$-Kj$&$Ki$&$Kj$ \\
\hline
$-k$&$1$&$k$&$-1$&$Kj$&$Ki$&$-Kj$&$-Ki$ \\
\hline
\hline
$Ki$&$-Kj$&$-Ki$&$Kj$&$1$&$-k$&$-1$&$k$ \\
\hline
$Kj$&$Ki$&$-Kj$&$-Ki$&$k$&$1$&$-k$&$-1$  \\
\hline
 $-Ki$&$Kj$&$Ki$&$-Kj$&$-1$&$k$&$1$&$-k$ \\ 
\hline
 $-Kj$&$-Ki$&$Kj$&$Ki$&$-k$&$-1$&$k$&$1$ \\ 
\hline
\end{tabular}
\end{center}

\caption{Cayley multiplication table for the co-quaternion group or dihedral D$_4$ that can be characterized by two parameters: $a = k, b = Ki, b^2 =1, a^2=-1$, with $ab=-ba=Kj$. Contrast with Table III.}
\end{table}

The next step is to two independent (each commutes with all of the other set) quaternions, which can be denoted by lower and upper case $(i,j,k)$ and $(I,J,K)$, along with the unit element and all bilinear products. Taking all sixteen with plus/minus signs forms the 32-element finite group Q$_{32}$. A half-way step is to include just one of the upper case, say $K$, to get a group of order-16. This element could also be regarded as an ordinary complex square root of unity so that we might call this the complex quaternion group, although it has been referred to as ``bi-quaternion," a term that Hamilton himself seems to have introduced (bi-quaternion could more properly have been kept for the full order-32 group with all multiplicative combinations of two independent quaternions) \cite{ref67}. Its Cayley table is shown in Table IV and its 15 elements placed in 1:1 correspondence with the Clifford cube and tetrahedron of Figs. 7 and 8. As a group, it is C$_2 \otimes$ Q$_8$ or $(I,Kk) \otimes \pm(I, i, j, k)$. (The pair $(1,K)$ would also give all 16 elements but do not form a C$_2$ group as do $(1, Kk)$.)

In terms of the extension from the seven point/line triangle/2-simplex to these 8-vertex/28-line cube or 15-point/35-line tetrahedron/3-simplex, the previous 3-binary is extended to a 4-binary $(txyz)$ and, geometrically, a new vertex (1000) or $K$ in a new dimension is introduced. In the cube, that vertex is connected to the previous seven $(0xyz)$ as mid-points and extended to vertices $(1xyz)$ whereas in the tetrahedron, the seven lines from vertex to the base Fano triangle introduce seven new midpoints, three of them face centers and one body center. Opposite vertices of the cube are of opposite sign: $\pm(K, Ki, Kj, Kk)$ and the center is the $\pm 1$ of quaternions and $ZZ$ generator of the example chosen. With the quaternions placed at midpoints of the edges at the lower corner, their negatives stand at the midpoints of the orthogonal face in keeping with their bi-vector nature noted earlier. In the tetrahedron, the new vertex $K$ introduces edge midpoints $(Ki, Kj, Kk)$, face centers the same with minus sign, and its conjugate $-K$ as body center, with $\pm 1$ and $ZZ$ the face center of the Fano triangle remaining in the base. Table II brings together all the alternative renderings of the 15 generators of the two-qubit system in terms of $O_i$, Pauli matrices, Dirac gamma matrices, binary, and quaternionic labels. Note the consistent pattern of each next simplex having a $(-k)$-like center with a string of 1's as its binary representation.

The 4-binary has a natural language in terms of space-time $(txyz)$ in physics but could equally be rendered in alternatives such as four colors \cite{ref61} or four acoustic notes to describe similar constructs of 15 basic objects and has been used \cite{ref46} in the context of a well-known combinatorics problem to be described below. In Table II, the association of this round bracket 4-binary with the bi-quaternions is fixed, the latter read off the former with the simple 1:1 association introduced in an above paragraph: $(0xyz) \leftrightarrow (kij)$, that is, $(0100) \leftrightarrow k$, etc. and conjugates such as $(0011) \leftrightarrow -k$. An initial 1 brings in $\pm K$ : $(1000) \leftrightarrow K, (1111) \leftrightarrow -K, (1010) \leftrightarrow Ki$, etc. Multiplication of quaternions corresponds to binary addition. However, since any $O_i$ denoted by its spinor and square bracket binary can be chosen as the center, there is no one to one link of them, the correspondence shown being for the specific choice of $ZZ$ as center and $\pm 1$.

\begin{table*}
\begin{center}
\begin{tabular}{|c|c|c|c|c|c|c|c||c|c|c|c|c|c|c|c|}
\hline
$1$&$k$&$-1$&$-k$ &$K$&$-K$&$Kk$&$-Kk$&$i$&$j$&$-i$&$-j$&$Ki$&$-Ki$&$Kj$&$-Kj$ \\
\hline
$k$&$-1$&$-k$&$1$&$Kk$&$-Kk$&$-K$&$K$&$j$&$-i$&$-j$&$i$&$Kj$&$-Kj$&$-Ki$&$Ki$ \\
\hline
$-1$&$-k$&$1$&$k$&$-K$&$K$&$-Kk$&$Kk$&$-i$&$-j$&$i$&$j$&$-Ki$&$Ki$&$-Kj$&$Kj$ \\
\hline
$-k$&$1$&$k$&$-1$&$-Kk$&$Kk$&$K$&$-K$&$-j$&$i$&$j$&$-i$&$-Kj$&$Kj$&$Ki$&$-Ki$ \\
\hline
$K$&$Kk$&$-K$&$-Kk$&$-1$&$1$&$-k$&$k$&$Ki$&$Kj$&$-Ki$&$-Kj$&$-i$&$i$&$-j$&$j$ \\
\hline
$-K$&$-Kk$&$K$&$Kk$&$1$&$-1$&$k$&$-k$&$-Ki$&$-Kj$&$Ki$&$Kj$&$i$&$-i$&$j$&$-j$ \\
\hline
$Kk$&$-K$&$-Kk$&$K$&$-k$&$k$&$1$&$-1$&$Kj$&$-Ki$&$-Kj$&$Ki$&$-j$&$j$&$i$&$-i$ \\
\hline
$-Kk$&$K$&$Kk$&$-K$&$k$&$-k$&$-1$&$1$&$-Kj$&$Ki$&$Kj$&$-Ki$&$j$&$-j$&$-i$&$i$ \\
\hline
\hline
$i$&$-j$&$-i$&$j$&$Ki$&$-Ki$&$-Kj$&$Kj$&$-1$&$k$&$1$&$-k$&$-K$&$K$&$Kk$&$-Kk$ \\
\hline
$j$&$i$&$-j$&$-i$&$Kj$&$-Kj$&$Ki$&$-Ki$&$-k$&$-1$&$k$&$1$&$-Kk$&$Kk$&$-K$&$K$ \\
\hline
$-i$&$j$&$i$&$-j$&$-Ki$&$Ki$&$Kj$&$-Kj$&$1$&$-k$&$-1$&$k$&$K$&$-K$&$-Kk$&$Kk$ \\
\hline
$-j$&$-i$&$j$&$i$&$-Kj$&$Kj$&$-Ki$&$Ki$&$k$&$1$&$-k$&$-1$&$Kk$&$-Kk$&$K$&$-K$ \\
\hline 
$Ki$&$-Kj$&$-Ki$&$Kj$&$-i$&$i$&$j$&$-j$&$-K$&$Kk$&$K$&$-Kk$&$1$&$-1$&$-k$&$k$ \\
\hline
$-Ki$&$Kj$&$Ki$&$-Kj$&$i$&$-i$&$-j$&$j$&$K$&$-Kk$&$-K$&$Kk$&$-1$&$1$&$k$&$-k$ \\
\hline
$Kj$&$Ki$&$-Kj$&$-Ki$&$-j$&$j$&$-i$&$i$&$-Kk$&$-K$&$Kk$&$K$&$k$&$-k$&$1$&$-1$ \\
\hline
$-Kj$&$-Ki$&$Kj$&$Ki$&$j$&$-j$&$i$&$-i$&$Kk$&$K$&$-Kk$&$-K$&$-k$&$k$&$-1$&$1$ \\
\hline
\end{tabular}
\end{center}
\caption{Cayley table for group of complex quaternions $\pm(1, i, j, k, K, Ki, Kj, Kk)$, an order-16 group isomorphic to C$_2 \otimes$ Q$_8$ and C$_2 \otimes$ D$_4$ with C$_2 = (1, Kk)$. An eight-element sub-group in the diagonal blocks is another alternative to Tables III and IV.}
\end{table*}
  
There are other order-8 sub-groups of the full order-16 complex quaternion group. One is the set, $\pm (I, k, K i, K j)$, forming the co-quaternion group D$_4$ with Cayley table shown in Table IV. In a standard minimal notation \cite{ref67} for D$_4$, it can be rendered in terms of two parameters as $(a=k, b=Ki)$. With the same labelling in Table II, the set of seven generators are $(ZZ, XX, YY, IY, ZX, YI, XZ)$ with $YY$ as center. Correspondingly, the physical system now is of two spins in a magnetic field in the $y$-direction with four magnetic coupling operators in the orthogonal $X-Z$ plane. And another is the set of eight elements $\pm(1, k, K, Kk)$ which is $(1, Kk) \otimes \pm(1, k)$ or C$_2 \otimes$ C$_4$. It corresponds for the example in Table II to the set $(ZZ, XX, YY, IX, ZY, XI, YZ)$ with $XX$ as center, that is, two spins now in a magnetic field in the $x$-direction and coupling terms in the orthogonal $Y-Z$ plane. Differing only in a renaming of the magnetic field in terms of $x, y, z$ direction, they share the same entanglement and other physics. All these sets show the same $I \leftrightarrow Z, X \leftrightarrow Y$ duality noted above. The full set of sixteen elements of a complex quaternion or the generators of SU(4) in Table II have a Cayley table shown in Table V and may be viewed as direct products of C$_2$ with the order-8 groups, whether Q$_8$ or D$_4$.

As noted, the quaternion labels placed on the vertices in Fig. 5 is arbitrary given the natural geometric symmetries of the triangle such as rotations through multiples of $\pi/3$. The independent placement of the qubit generators is also arbitrary as, further, is their correspondence to the quaternions. Given that any of the 15 $O_i$ of them can serve as center and define $X$-states of the Fano sub-group, there are as many choices but entanglement properties differ, only the nine $\tau_i \sigma_j$ involving both qubits accommodating quantum entanglement. Their set of seven generators has two single spin generators, the $\tau_i$ and $\sigma_j$, and five two-spin operators. All square to unity. For the six single-spin centers, the similar set is composed of the three single operators of the other spin and the three products of them with the center. With these generators, a multiplicative sign or constant such as $\pm i$ is irrelevant and, as observed, the $-i\vec \sigma$ satisfy all the multiplication rules of quaternions including that they square to $-1$. Turning to the order-8 groups, the quaternion group Q$_8$ has six $-1$ along the diagonal in its Cayley Table III, while the co-quaternion group in Table IV that is isomorphic to the dihedral D$_4$ has two, and the C$_2 \times $ C$_4$ has four (the upper left 8 $\times$ 8 block of Table V). Thus, while all three of them can be associated with $X$-states, it is the co-quaternion which matches best the entangled class in having two of the seven square to -1. Those could be set to match in the set with center $\tau_i \sigma_j$ the two, $\tau_i$ and $\sigma_j$, with $-i$ factors, thus squaring to -1. 

Thus, the choice made in Fig. 5, 7 and 8, and Table II puts $\pm i$ as the two $IZ$ and $ZI$ with a corresponding co-quaternion $\pm(1, i, Kj, Kk)$, a set trivially different from the one displayed in Table IV. For this choice, all rows of Table II can be retained unchanged, especially the binary and qubit generators, but in the last row, multiply entries involving $(j,k)$ by $K$ with $K^2 =-1$. In Figs. 5, 7 and 8, similarly multiply $(j,k)$ by $K$, other points left unchanged. On the other hand, had we chosen the quaternion $\pm(1, k, Ki, Kj)$, the entries in Table II could be left unchanged but this would correspond to the $X$-state set $(ZZ, XX, YY, IY, ZX, YI, XZ)$ with $YY$ as center and in slight conflict with the convention in physics of choosing $z$ as the magnetic field direction or quantization axis. As stated, with no unique correspondence between the three labelling systems, binary, quaternion, and qubit generators, it is partly convention and partly aesthetics dictating the choice made in Table II and in our figures. The choice made in Table II is to tie square bracket 4-binary to qubit generators and round bracket to bi-quaternions, with $ZZ$ as center of the former matched to the $\pm 1$ center of quaternions.   

An interesting connection can also be established with octonions, the only other division algebra besides reals, complex numbers, and quaternions \cite{ref55}. With seven independent square roots of -1, they can also be laid on the seven points of Fig. 5 but with crucial differences. With a cyclic triplet $(pqr)$ replacing $(-i, -j, -k)$ in order, and the seventh square root $s$ placed at the center, these seven independent imaginaries $(i, j, k, p, q, r, s)$ have seven cyclic lines on the triangle $(irj), (jpk), (kqi), (psi), (qsj), (rsk), (pqr)$. All lines are now arrowed with the same circulation sense for the edges and the circle unlike in Fig. 5 that has them opposite, and medians now also arrowed from midpoint to vertex. This turns out to be crucial because octonionic multiplication is no longer associative as with the previous three division algebras, that opposite circulation a necessary feature of quaternions in Fig. 5 in contrast to octonions. Interestingly, the seven quaternionic triplets can also be depicted on the cube in Fig. 8 with the unit element at the lower left corner, $(ijk)$ on the three connected vertices to it, and $(pqr)$ at their opposite corners and $s$ at the body center. An alternative placement of seven at seven corners of the cube is in \cite{ref55}.

This aspect of counting how many -1 occur along the diagonal of a Cayley table that arises naturally in our discussion with qubit generators also has a bearing on further extension. There are no more division algebras beyond octonions to place on all points after the 2-simplex of Fig. 5. In the 3-simplex tetrahedron of Fig. 7, $\pm(i, j, k, K, Ki, Kj, Kk)$, eight of them square to -1. 15 square roots of -1, placed at each of the points represent what are called ``sedenions" \cite{ref67a}. Products of pairs of them as 35 triplets can be specified and, again as with octonions, the products are not commutative or associative. That loss of associativity in multiplication precludes, of course, matrix representation and physics has seen little use of octonions or sedenions (see, however, \cite{ref67b,ref67c} and octonions do allow what is termed ``limited associativity" \cite{ref55}). But, from the 35 triplets we have discussed for the tetrahedron in either quaternionic or spinor language and a Cayley table such as Table V, one can build a Cayley table for sedenions as in \cite{ref67a}. Indeed, a more symmetric arrangement than in \cite{ref67a} is to group the 35 triplets into seven columns of 5 rows each with all 15 elements occurring once and only once in each column for the Kirkman arrangement of schoolgirls discussed below in Sec. IV. Note that a proper Cayley table requires the 15 square roots with both plus/minus signs along with $\pm 1$, and is a group of order 32. As an alternative to the Cayley-Dickson construction in \cite{ref67a}, the higher $q$-qubit simplexes provide another route to constructing these hypercomplex numbers of sedenions and beyond and associating them with finite projective geometries.           

The correspondence to quaternions and higher string binaries for more than two qubits proceeds naturally. Each further qubit in a $q$-qubit sequence introduces a new initial entry of 1 in the string with a new independent $K$-like entry and new vertex in the next simplex, the previous simplex's points assigned an initial 0. $-K$ appears with $(111 \ldots )$ as the new simplex's body center, just as mid-point, face center, tetrahedron's body center, did for 1-, 2-, 3-simplex of 1-, 2-, 3-qubit systems, respectively. This gives a geometric realization in simplexes of $X$-states of $q$-qubits. A somewhat different approach was adopted in \cite{ref67d} that followed the Cayley-Dickson $2^N$-dimensional algebra with imaginary units $e_a, 1 \leq a \leq 2^N-1$ encoded in PG($N -1$, 2), triads of points with $e_a e_b = \pm e_c$ as lines. Binomial configurations $C_N$ are then identified with octonions for $N=3$, sedenions for $N=4$, and higher $2^N$--nions. Again, the PG($N-1$, 2) is a $(N-1)$-dimensional projective space over Galois field GF(2) as noted at the end of Sec. III D and the $C_N$ are isomorphic to Grassmannian G$_2(N+1)$. While \cite{ref65b} noted that there is no ``neat picture" for these higher 2$^N$-nions, the $N$-binary string and simplex schemes discussed above provide such a unified picture of all of them and associate with the $X$-states (or, equivalently, pure states) of $N$-qubits. And, in the correspondence to quaternions, at each step one new independent $K$-like imaginary unit is added to get to the simplex of the next higher dimension.    

\section{Geometric View}

In the last twenty years, two lines of exploration with very different starting points and motivations have come together in studying symmetries of systems with a finite number of quantum spins. One that we have discussed so far in Sec. III started with a concrete physical problem in nmr of two coupled spins \cite{ref7}. The observation that sub-group symmetries in such a system simplify the construction of the evolution operator led in subsequent work to generalization and systematics of SU(2) $\times$ U(1) $\times$ SU(2) and other sub-groups of SU(4) \cite{ref8,ref34,ref46}. These studies of such continuous Lie groups and Lie algebras were later connected to quaternions and their discrete group symmetries and to projective geometries PG with nice geometric figures of triangles, tetrahedrons and higher simplexes to describe quantum states and operators, along with geometric manifolds that generalize the Bloch Sphere of a single spin \cite{ref46,ref50}. At the same time and in the same period, geometers have investigated objects entirely within geometric algebra and arrived at a similar picture \cite{ref49,ref52,ref53,ref56}. This section will deal with that approach.

Three such early works \cite{ref13,ref53,ref68} approached the ``geometry of entanglement" for two qubits by considering a six-dimensional real metric vector space V with a non-degenerate quadratic form Q: V $\rightarrow$ R. Whereas earlier sections identifying SO(6) pointed to a corresponding six-dimensional space of real rotations, these geometers used a generalization of vector algebra to metric vector spaces in geometric algebra. An antisymmetric product in V is called a r-blade and in the exterior algebra $\Lambda$V, a geometric product of a vector and an r-blade is defined. The number of vectors r in an r-blade is called its grade. Spinors are defined as left-ideals in a three-dimensional vector space G(3), its even parity elements being the quaternions denoted G$^{+}(3)$. Similarly, for two qubits, a space G(6) is defined with an orthonormal basis in R$^6$ of two triplets ($e_i, f_i$) \cite{ref13}. Bilinear combinations, bivectors, of $\Lambda$V are 

\begin{equation} 
G_{10} = e_2e_3, G_{01} = f_2f_3, G_{ij} = e_i f_j (-1)^{\delta_{ij}},
\label{eqn15}
\end{equation}
and similar cyclic combinations. These 15 $G$'s can be placed in 1:1 correspondence with the $O_i$ introduced earlier and are also shown in Table II. They can also be conveniently depicted as a hexagon in Fig. 9 \cite{ref13}. When one of the subscripts of $G$ is zero, they correspond to single qubit operators within $e_i$ and $f_i$, and they stand on the left and right of the hexagon. Only the nine $G_{ij}$ with $i, j =1-3$, that is two-spin operators, as centers support entanglement as per earlier remarks. It is these $G$'s that are the Dirac $(\gamma_i, A_i, \alpha_i)$.  

\begin{figure}
\centering
\scalebox{1.5}{\includegraphics[width=3.5in]{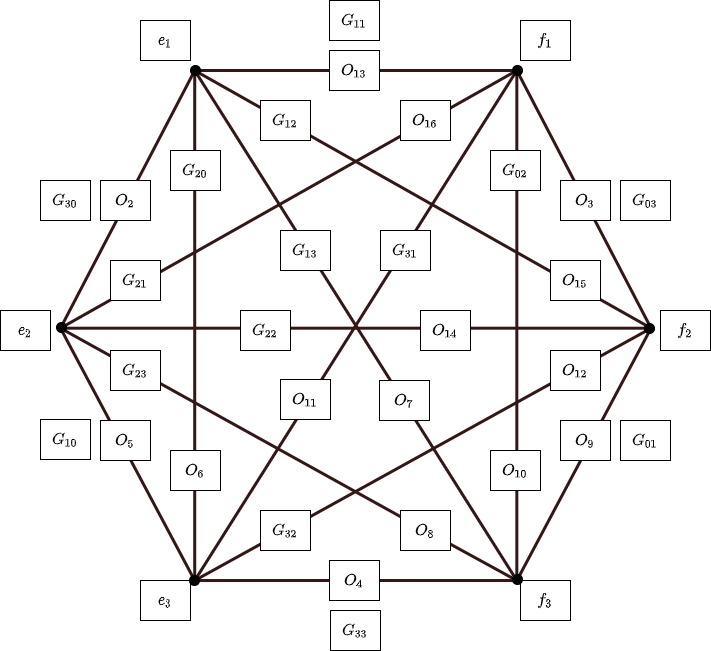}}
\caption{Geometric algebra of six-dimensional space of two vectors $e_i, f_i$ for each qubit and bivectors $G_{ij}$ and their correspondence to the $O_i$ generators of the qubit-qubit system. Only the nine $G_{ij}$ links admit quantum entanglement, not the $G_{0i}$ and $G_{i0}$ associated with the individual qubits. Adapted from \cite{ref13}.}
\end{figure}

Pure states of two-qubits, characterized by seven real parameters, are described in S$^7$. The three-dimensional projective space P(V) = PG(3, 2) has an underlying vector space of dimension four. These are referred to as the boundary and bulk, respectively. Going back to the work of Pl\"{u}cker, Klein, and Grassmann, lines of the projective space can be parametrized in terms of points in four dimensions. There are 35 lines and 15 points of PG(3, 2) as described earlier in the tetrahedron of Fig. 7. It is useful to fiber S$^7$ over a one-dimensional quaternionic projective space HP$^1 \sim$ S$^4$ by a second Hopf fibration $\pi$ : S$^7$ $\rightarrow$ S$^4$ with an SU(2) $\sim$ S$^3$ fiber. A  natural metric, the Mannoury-Fubini-Study metric, is induced which is the standard metric on S$^4$ expressed in stereographically projected coordinates. The geodesic distance with respect to this metric provides a natural object for quantifying entanglement of the qubits according to the prescription that entanglement is the geodesic distance to the nearest separable state \cite{ref68,ref69}. Entanglement resides, therefore, in the twisting of the bundle between the base S$^4$ and fiber S$^3$. (This usage of S$^4$ should not be confused with the one in Fig. 6 for the Desargues sub-group in Sec. III B.) Using this, geometric meaning is given to the standard Schmidt decomposition that is familiar in quantum information. The Schmidt states are the nearest and the furthest separable states lying on, or the ones obtained by parallel transport along, the geodesic passing through the entangled state \cite{ref68}. That geodesic distance is expressible \cite{ref68} in terms of the ``concurrence" which quantifies entanglement in quantum information \cite{ref70}. Another natural way of saying this is through a connection on the bundle. Sec. IV of \cite{ref68} identifies it as the instanton connection familiar in quantum field theory where it describes tunneling between different continua. Further, a Sp(1) $\sim$ SU(2) gauge degree of freedom of the second Hopf fibration gives an important geometric interpretation of local transformation in the second subsystem not changing the entanglement properties of the whole. Those can only be affected by a global unitary U(4).

Other geometric objects are a generalized Klein quadratic, denoted as W$_3$(2). It is a hyperbolic quadric in W(3, 2), the symplectic polar space of rank 2 and order 2, and the space of totally isotropic subspaces of PG(3, 2) with respect to a symplectic form. In PG(3, 2) with 15 points and 35 lines, 7 lines are incident on each point, three of them isotropic and 4 non-isotropic (commuting and cyclic, respectively, as referred to in Lie algebraic terminology of Sec. III). W(3, 2) is a self-dual object of 15 points and 15 lines, called a ``doily" or a Cremona-Richmond configuration \cite{ref49,ref71} (Fig. 1 of \cite{ref52}). It is the smallest generalized quadrangle. A decomposition into a 10-point/line Petersen graph and five leftover points, none of them collinear, called an ovoid is also shown. Also, two distinct points of W(3, 2) are said to be orthogonal if joined by a line, again what was termed commuting in our earlier Lie algebra language. W(3, 2) has three kinds of hyperplanes: 1) called perp-set, which is a set of points collinear with a given point; there are 15 such and each a Fano Plane or also called pencil, 2) a grid of 9 points on 6 lines called a Pappus configuration and Mermin square in applications in quantum information such as proofs of the Kochen-Specter theorem \cite{ref72}; there are 10 such, 3) an ovoid which are 5 points with exactly one point in common with every line; there are six such. The dual of an ovoid is called a ``spread" \cite{ref53}. A so-called Veldkamp space of the doily V(W(3, 2)), which is a parabolic quadric and isomorphic to PG(4, 2) with 31 points (and 155 lines) of which 15 are generated by single-point perp-sets, 10 by grids, and 6 by ovoids, has also been discussed \cite{ref71}. And a Mermin pentagram has been discussed in \cite{ref72a}. 
   
Besides entanglement, the quadric W(3, 2) and the correspondence between two-qubit observables on the boundary and three-qubit ones in the bulk have been discussed for other purposes in quantum information such as the generation of error-correcting and stabilizer codes. The ovoid of five points represents five codewords encoding messages of the boundary \cite{ref52}. It also gives a way of getting the maximum number of MUBs (mutually unibiased base) in a finite-dimensional Hilbert space, otherwise a difficult problem that is however important in quantum information \cite{ref56}. A group and graph theoretic approach to MUBs has also been considered in terms of Cayley graphs \cite{ref73}. The basis group of a set of MUBs of a $d$-dimensional Hilbert space is defined by a sub-group of U($d$) generated by unitary matrices associated with the bases. The edges of the Cayley graph that captures this structure form a completely connected subgraph called a ``clique" \cite{ref74}. This links the search for MUBs to the representation theory of finite groups \cite{ref73}.

The geometric literature has also extended beyond two-qubits to $q$-qubits. This discussion has again been for pure states rather than for more general mixed states that are of more interest in the field of quantum information. The starting manifold of $q$-qubits is S$^{2^{q+1} -1}$ and Hopf fibration $\pi$ : S$^{2^{q+1} -1} \rightarrow$ S$^{2^q}$ with fiber S$^{2^{q}-1}$ \cite{ref75}. The projective geometry is now PG(2$q$-1, 2) and Klein quadric W(2$q$-1, 2). The roles of W(5, 2) for three qubits and W(7, 2) for four has also been discussed recently \cite{ref75a}. There are now $2^{2q-1}$ points not orthogonal to a given point instead of the eight non-zeroes in Table I of qubit commutators. Thus, PG(2$q$-1, 2) is cut into $2^q +1$ disjoint fibers, each containing $2^q -1$ points. For three-qubits, this amounts to 63 = 9 $\times $ 7 in place of the 15 = 5 $\times$ 3 for qubit-qubit. An interesting connection to a hoary mathematical problem called the Kirkman Schoolgirls problem that is associated with the latter decomposition is worth noting \cite{ref51}, since it influenced several developments in finite projective geometries \cite{ref76,ref77,ref78}, and design theory \cite{ref18,ref20,ref79,ref80,ref81}.

That introduces another branch of mathematics within its area of combinatorics. Going back to a recreational problem of over 175 years ago \cite{ref76} that has since been known as Kirkman's Schoolgirls problem, mathematicians have studied it as ``triple systems" within ``design theory." In particular, ``balanced incomplete block (BIB)" designs and ``Steiner triple systems" were related to finite projective geometries by mathematical statisticians, notably R. A. Fisher \cite{ref18,ref20,ref51,ref80,ref82}. A number $v$ of ``varieties" are assigned to ``blocks" $b$ with incidence relations to provide $(v, b, r, k, \lambda)$ designs. The symbols $v$ are assigned to blocks $b$ with $k$ in each, and each symbol to occur in $r$ different blocks with every pair of symbols to occur together in $\lambda$ blocks. The case of $k =3$ is designated a triple system and $\lambda = 1$ (no repeats) a Steiner system. The two conditions together define Steiner triple systems that have been extensively studied and exist for all $ v = 1$, or 3 mod 6. Since BIBs must satisfy $vr = bk, \lambda (v-1) = r(k-1)$, such a Steiner triple is fixed by the single parameter $v$ and denoted 2-($v$, 3, 1). Apart from the trivial $ v =3, b =1$, the next is $v=7, b=7$, an example of what is dubbed "symmetric" design. The binomial configuration $C_3$ associated with octonions that was discussed earlier has been recognized as isomorphic to a so-called ``Pasch" configuration and used for classifying Steiner triple systems \cite{ref67d}.

With $v$ taken as points and $b$ as lines, the incidence relation of projective geometry that three points lie on every line provides a connection to finite projective geometry. This recognition by Fisher and collaborators was very fruitful for the field of design theory which was born of those origins \cite{ref51,ref82}. Apart from the trivial single line design 2-(3, 3, 1) of $v=3$ points and PG(1, 2), the simplest next example of 2-(7, 3, 1) design is isomorphic to PG(2, 2) of the Fano Plane. Another design 2-(15, 3, 1) with $v=15, b=35$ is PG(3, 2) and it is this connection through the numbers 7 and 15 that led to mapping two-qubit problems to finite projective geometry and design theory \cite{ref46,ref50,ref51}. 7 points on 7 lines is the Fano triangle and 15 points on 35 lines the tetrahedron discussed in earlier sections. Kirkman's schoolgirls problem was to have 15 schoolgirls march 3 abreast to school every day of a 7-day week with no pairs of girls repeated in a row, and the 35 such rows can be drawn from the tetrahedron to provide such a marching order \cite{ref50,ref83}. The choice of three abreast in a recreational problem was a prescient and happy anticipation to three being the number of operators involved at a vertex (of Feynman, angular momentum coupling, many-body perturbation theory, etc.) and in a Lie commutator in quantum physics applications a century later. 

\begin{table}
\begin{center}
\begin{tabular}{|c|c|c|c|c|c|}

\hline
$q$&$n=2q-1$&$v$&$b$&$r$& -  \\ 
\hline 
$1/2$&$0$&$1$&$0$&$0$& PG(0, 2) \\
\hline
$1$&$1$&$3$&$1$&$1$& PG(1, 2) \\
\hline
$3/2$&$2$&$7$&$7$&$3$& PG(2, 2) \\
\hline
$2$&$3$&$15$&$35$&$7$& PG(3, 2) \\
\hline
$5/2$&$4$&$31$&$155$&$15$& PG(4, 2)  \\
\hline
 $3$&$5$&$63$&$641$&$31$& PG(5, 2) \\ 
\hline
 $7/2$&6&$127$&$2667$&$63$& PG(6, 2) \\ 
\hline
\end{tabular}
\end{center}
\caption{Triple system designs with $v$(arieties), $b$(locks), $r$(anks) and corresponding projective geometry PG. Integer values of $q$ represent number of qubits in correspondence.}
\end{table}

The connections apply to higher $q$-qubits as well, with 

\begin{equation}
v= 2^{2q} -1, b=(2^{2q} -1)(2^{2q-1} -1)/3, r= 2^{2q-1} -1,
\label{eqn16}
\end{equation}
and the geometry PG(2$q$ -1, 2) as illustrated in Table VI. While $k=3$ triplets are directly related to projective geometries as we have discussed and to quantum commutators when two operators uniquely fix the third (physics abounds in triplets such as vertices in angular momentum coupling or Feynman diagrams), the Kirkman problem can itself be generalized for other values and described in terms of PG($n, m$) with $m=k-1, n = 2q -1$, and 

\begin{equation}
v_n = mv_{n-1} + 1 = \frac{m^{n+1} - 1}{m-1}, b = \frac{(m^n -1)(m^{n+1} -1)}{(m -1)^2(m+1)}, r =\frac{m^n -1}{m-1}.
\label{eqn17}
\end{equation}
Such generalizations have found application in recreational examples of golfers in rounds of four or more that followed Kirkman's schoolgirl triplets \cite{ref83}.
The dimension of PG($n, m$) is shown in Table VII for place values of $m$ beyond the $m=k-1=2$ that has occurred throughout this paper as relevant to physics. For $m=1$, all entries being powers of 1, the column is simply $n+1$ and corresponds to pairs ($k=2$) instead of triplets. The $m=2, k=3$ column has the entries $1, 3, 7, \ldots $ for triplets discussed so far of PG($n, 2$). The next column of $m=3$ are quartet arrangements with dimension $(3^{n+1} -1)/2$. Along rows at fixed $n$ are the sequences 1, 11, 111, etc., with place value $m$, and thus, $1, m+1, m^2+m+1, \ldots, [m^{n+1}-1] /(m-1), \ldots$.

\begin{table}
\begin{center}
\begin{tabular}{|c||c|c|c|c|}

\hline
$n/m$&$1$&$2$&$3$&$4$ \\ 
\hline
-&pairs&triplets&quartets&quintets \\
\hline
\hline 
$0$&$1$&$1$&$1$&$1$  \\
\hline
$1$&$2$&$3$&$4$&$5$ \\
\hline
$2$&$3$&$7$&$13$&$21$  \\
\hline
$3$&$4$&$15$&$40$&$85$ \\
\hline
$4$&$5$&$31$&$121$&$341$   \\
\hline
\end{tabular}
\end{center}

\caption{ Dimension of PG($n, m$)}
\end{table}

\section{Higher dimensional spins}
Increasingly, spins larger than 1/2 are being explored for applications in quantum information \cite{ref84,ref85,ref86,ref87,ref88,ref89,ref90,ref91,ref92,ref93}. As is well known, entanglement measures such as concurrence \cite{ref70}, and negativity of the partial transpose \cite{ref94}, fail when both $d$ and $D$ in a qudit-quDit system are larger than 2 \cite{ref95}. A qutrit, with symmetry group SU(3) and symmetry algebra su(3), is characterized by 8 parameters, a general qudit of $d$-dimensions by su($d$) with $(d^2-1)$ parameters. While various geometric extensions of the qubit Bloch sphere \cite{ref96} have been advanced with corresponding Bloch vectors for a qudit \cite{ref97,ref98,ref99}, it is still difficult to visualize a single qutrit \cite{ref93,ref100} and there is no satisfactory way to view higher-$d$ state space. A recent work advances for this purpose projecting state space onto measurements of observables with illustrations for photonic qutrits that use three spatial modes of the electromagnetic field \cite{ref93}. Another recent picture of qutrits is in \cite{ref100a}; see also \cite{ref96,ref100b}. And, in earlier quantum physics literature, such a description of higher spin $j$ in terms of observables is in \cite{ref101,ref102} and a generalization of the Bloch equation as in Eq.~(\ref{eqn10}) and Eq.~(\ref{eqn12}) in terms of multipoles for spin-$j$ in Eq.(7.37) of \cite{ref2}.    

The 3 $\times$ 3 density matrix of a qutrit has two real parameters on the diagonal and three complex off-diagonal entries. An $X$-state of a qutrit that may be denoted SU$^{(X)}$(3) has only one off-diagonal for a total of four parameters. A central, real one-dimensional space is decoupled from the two-dimensional space surrounding it, making SU$^{(X)}$(3) of SU(2) $\times$ U(1) symmetry. In extending to multiple qutrits, the symmetry structure generalizes the case of multiple qudits. The 9 $\times$ 9 matrix of a qutrit-qutrit has in general 80 parameters characterizing it and \cite{ref67c} discusses it in terms of a two-qutrit Pauli group. But, again $X$-states involve a smaller number of parameters, only 16, consisting of 8 real diagonal entries and four off-diagonal complex elements. It can be seen in relation to the SU$^{(X)}$(3) of a single qutrit as repeating three copies (instead of two copies noted earlier for similar qubit extension) with U(1)s in between and at the ends \cite{ref63}. The 9-dimensional space has a central real one by itself and surrounded by four decoupled two-dimensional spaces.This structure of $X$-states says that in any even dimension, the system's density matrix may be viewed as $d/2$ independent U(2)s with one overall trace condition, whereas in odd dimension, there are $(d-1)/2$ such U(2)s and a central U(1) with again the trace condition. Similar and straightforward extensions to general qudit-quDit systems have been discussed in \cite{ref63} with an enumeration of the parameter space involved. There remain many more connections to projective geometries and Clifford algebra to be explored.

\section{Acknowledgments}
I thank J. P. Marceaux for many discussions and for preparing several of the figures. I also thank the Alexander von Humboldt Stiftung for support and Prof. Gernot Alber for discussions and hospitality at Technische Universit\"{a}t, Darmstadt.

\end{document}